%% file: HT-ITW_ev6.tex
\documentclass[onecolumn]{IEEEtran}
\input{preamble}

\usepackage{cite}
\usepackage{amsmath}
\usepackage{amssymb}
\usepackage{bbm}
\usepackage{color}

\usepackage[ colorlinks = true,
       linkcolor = blue,
       urlcolor = blue,
       citecolor = red,
       anchorcolor = green,]{hyperref}
\allowdisplaybreaks[1]

\begin{document}

\title{Strong Converse for Hypothesis Testing Against Independence over a Two-Hop Network}
\author{\IEEEauthorblockN{Daming Cao, Lin Zhou and Vincent Y. F. Tan} 
\thanks{Daming Cao is with School of Information Science and Engineering, 
Southeast University, China (Email: dmcao@seu.edu.cn). Lin Zhou and Vincent Y. F. Tan are with the Department of Electrical and Computer Engineering, National University of Singapore (Emails: lzhou@u.nus.edu, vtan@nus.edu.sg). Vincent Y. F. Tan is also with the Department of Mathematics, National University of Singapore.}
}

% make the title area
\maketitle

\begin{abstract}
By proving a strong converse, we strengthen the weak converse result by Salehkalaibar, Wigger and Wang (2017) concerning hypothesis testing against independence over a two-hop network with communication constraints. Our  proof follows by judiciously combining two recently proposed techniques for proving strong converse theorems, namely the strong converse technique via reverse hypercontractivity by Liu, van Handel, and Verd\'u (2017) and the strong converse technique by Tyagi and Watanabe~(2018), in which the authors used a change-of-measure technique and replaced hard Markov constraints with soft information costs. The techniques  used in our paper can also be applied to prove strong converse theorems for other multiterminal hypothesis testing against independence problems.
%such as the interactive hypothesis testing problem by Xiang and Kim (ISIT 2012), the cascaded hypothesis testing problem by Zhao and Lai (ITW 2015) and the hypothesis testing over the Gray-Wyner network by Wigger and Timo (SPCOM 2016).
\end{abstract}

\IEEEpeerreviewmaketitle

\section{Introduction}
Motivated by situations where the source sequence is \emph{not} available directly and can only be obtained through limited communication with the data collector, Ahlswede and Csisz{\'a}r \cite{ahlswede1986hypothesis} proposed the problem of hypothesis testing with a communication constraint. In the setting of \cite{ahlswede1986hypothesis}, there is one encoder and one decoder. The encoder has access to one source sequence $X^n$ and transmits a compressed version of it to the decoder at a limited rate. Given the compressed version and the available source sequence $Y^n$ (side information), the decoder knows that the pair of sequences $(X^n,Y^n)$ is generated i.i.d.\ from one of the two distributions and needs to determine which distribution the pair of sequences is generated from. The goal in this problem is to study the tradeoff between the compression rate and the exponent of the type-II error probability under the constraint that the type-I error probability is either vanishing or non-vanishing. For the special case of testing against independence, Ahlswede and Csisz{\'a}r provided an exact characterization of the rate-exponent tradeoff. They also derived the so-called strong converse theorem for the problem. This states that the rate-exponent tradeoff cannot be improved even when one is allowed a non-vanishing type-I error probability. However, the characterization the rate-exponent tradeoff for the general case (even in the absence of a strong converse) remains open till date.

Subsequently, the work of Ahlswede and Csisz{\'a}r was generalized to the distributed setting by Han in \cite{han1987hypothesis} who considered hypothesis testing over a Slepian-Wolf network. In this setting, there are two encoders, each of which observes one source sequence and transmits a compressed version of the source to the decoder. The decoder then performs a hypothesis test given these two compression indices. The goal in this problem is to study the tradeoff between the coding rates and the exponent of type-II error probability, under the constraint that the type-I error probability is either vanishing or non-vanishing. Han derived an inner bound to the rate-exponent region. For the special case of zero-rate communication, Shalaby and Papamarcou~\cite{shalaby1992multiterminal}  applied the blowing-up lemma~\cite{csiszar2011information} judiciously to derive  the exact rate-exponent region and a strong converse theorem. Further generalizations of the work of Ahlswede and Csisz{\'a}r can be categorized into two classes: non-interactive models where encoders do not communicate with one another~\cite{amari1998statistical,tian2008successive,wigger2016testing,zhao2016distributed} and the interactive models where encoders do communicate~\cite{xiang2012interactive,zhao2018distributed}. 
% Often, the performance of these problem is measured by the type-II error exponent under the constraints that the communication rates are bounded and the type-I error probability is vanishing. This criterion facilitates a weak converse proof relying on data process inequality. The strong converse theorem claims that the type-II error exponent can not be improved even the type-I error probability is non-vanishing. 

\begin{figure}[t]
\centering
\setlength{\unitlength}{0.5cm}
\scalebox{0.9}{
\begin{picture}(12,6)
\linethickness{1pt}
\put(2.1,0){\makebox(0,0){$X^n$}}
\put(7.1,0){\makebox(0,0){$Y^n$}}
\put(12.1,0){\makebox(0,0){$Z^n$}}
%put encoders and decoder
\put(11,2){\framebox(2,2)}
\put(6,2){\framebox(2,2)}
\put(1,2){\framebox(2,2)}
%put names for encoders and decoder
\put(6.7,2.8){\makebox{$f_2$}}
\put(1.7,2.8){\makebox{$f_1$}}
\put(11.7,2.8){\makebox{$g$}}
%put vectors to encoders and decoder
\put(2,0.5){\vector(0,1){1.5}}
\put(7,0.5){\vector(0,1){1.5}}
\put(12,0.5){\vector(0,1){1.5}}
%put vectors between encoders and decoder
\put(8,3){\vector(1,0){3}}
\put(3,3){\vector(1,0){3}}
\put(7,4){\vector(0,1)2}
\put(9.5,3.5){\makebox(0,0){$M_2$}}
\put(4.5,3.5){\makebox(0,0){$M_1$}}
\put(12,4){\vector(0,1)2}
\put(7.1,6.5){\makebox(0,0){$\hatH_Y$}}
\put(12.1,6.5){\makebox(0,0){$\hatH_Z$}}
\end{picture}}
\caption{System model for hypothesis testing over a two-hop network}
\label{systemmodel}
\end{figure}
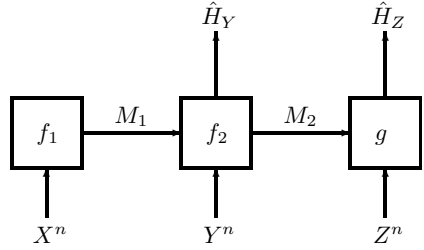

We revisit one such interactive model as shown in Figure \ref{systemmodel}. This problem was considered by Salehkalaibar, Wigger and Wang in~\cite{salehkalaibar2017hypothesis} and we term the problem as {\em hypothesis testing over a two-hop network}. The main task in this problem is to construct two hypothesis tests between two joint distributions $P_{XYZ}$ and $Q_{XYZ}$. One of these two distributions governs the law of $(X^n,Y^n,Z^n)$ where each copy $(X_i, Y_i, Z_i)$ is generated independently either from $P_{XYZ}$ and $Q_{XYZ}$. As shown in Figure~\ref{systemmodel}, the first terminal has knowledge of a source sequence $X^n$ and sends an index $M_1$ to the second terminal, which we call the relay; the relay, given side information $Y^n$ and compressed index $M_1$, makes a guess of the hypothesis $\hatH_Y$ and sends another index $M_2$ to the third terminal; the third terminal makes another  guess of the hypothesis  $\hatH_Z$ based on $M_2$ and its own side information $Z^n$. The authors in \cite{salehkalaibar2017hypothesis} derived an inner bound for the rate-exponent region and showed that the bound is tight for several special cases, including the case of testing against independence in which $Q_{XYZ}=P_XP_YP_Z$. However, even in this simpler case of testing against independence, which is our main  concern in this paper, the authors in \cite{salehkalaibar2017hypothesis} \emph{only} established a \emph{weak} converse.

In this paper, we strengthen the result by Salehkalaibar, Wigger and Wang in \cite{salehkalaibar2017hypothesis} by deriving a \emph{strong} converse for the case of testing against independence. Our proof follows by \emph{judiciously} combining two recently proposed strong converse techniques by Liu \emph{et al.} in \cite{liu2017beyond} and by Tyagi and Watanabe in \cite{tyagi2018strong}. In \cite{liu2017beyond}, the authors proposed a framework to prove strong converse theorems based on functional inequalities and   reverse hypercontractivity of Markov semigroups. In particular, they  applied their framework to derive strong converse theorems for a collection of problems including  the  hypothesis testing with communication constraints problem  in \cite{ahlswede1986hypothesis}. In \cite{tyagi2018strong}, the authors proposed another framework for strong converse proofs, where they used a change-of-measure technique and replaced hard Markov constraints with soft information costs. They also leveraged variational formulas for various information-theoretic quantities; these formulas were  introduced by Oohama in \cite{oohama2016wynerziv,oohama2015exponent}.

\subsubsection*{Notation}
\label{sec:notation}%\red{remove all notation that are not used. I think type $\hatP_{x^n}$ is not used!!} \blue{type $\hatP_{x^n}$ is used in the definition of the typical set $\calT_{n}(P(Y))$ in \eqref{def:calT}}
Random variables and their realizations are in upper (e.g.,\ $X$) and lower case (e.g.,\ $x$) respectively. All sets are denoted in calligraphic font (e.g.,\ $\calX$). We use $\calX^{\rmc}$ to denote the complement of $\calX$. Let $X^n:=(X_1,\ldots,X_n)$ be a random vector of length $n$ and $x^n$ its realization. Given any $x^n$, we use $\hatP_{x^n}$ to denote its type (empirical distribution). All logarithms are base $e$. We use $\bbR_+$ and $\bbN$ to denote the set of non-negative real numbers and  natural numbers respectively. Given any positive integer $a\in\bbN$, we use $[a]$ to denote $\{1,\cdots, a\}$. We use $1\{\cdot\}$ to denote the indicator function and use standard asymptotic notation such as $O(\cdot)$. The set of all probability distributions on a finite set $\calX$ is denoted as $\calP(\calX)$. Given any two random variables $(X,Y)$ and any realization of $x$, we use $P_{Y|x}(\cdot)$ to denote the conditional distribution $P_{Y|X}(\cdot|x)$. Given a distribution $P\in\calP(\calX)$ and a function $f:\calX\to\calR$, we use $P(f)$ to denote $\bbE_{P}[f(X)]$. For information-theoretic quantities, we follow \cite{cover2012elements}. In particular, when the joint distribution of $(X,Y)$ is $P_{XY}\in\calP(\calX\times\calY)$, we use $I_{P_{XY}}(X;Y)$ and $I(X;Y)$ interchangeably. Throughout the paper, for ease of notation, we drop the subscript for distributions when there is no confusion. For example, when the joint distribution of $(X,Y,Z)$ is $P_{XYZ}$, we use $I_{P}(X;Y|Z)$ and $I_{P_{XYZ}}(X;Y|Z)$ interchangeably. For ease of notation, for any $(p,q)\in [0,1]^2$, let $D_\rmb(p\|q)$ denote the binary divergence function, i.e., $D_\rmb(p\|q)=p\log (p/q)+(1-p)\log((1-p)/(1-q))$. %Finally, for any vector $$, we use $\|X^n\|$ to denote the $L_1$ norm, i.e., $\|X^n\|=\sum_{i\in[n]} |x_i|$.

\section{Problem Formulation and Existing Results}
\subsection{Problem Formulation}
Fix a joint distribution $P_{XYZ}\in\calP(\calX\times\calY\times\calZ)$ satisfying the Markov chain $X-Y-Z$, i.e., 
\begin{align}
P_{XYZ}(x,y,z)=P_{XY}(x,y)P_{Z|Y}(z|y)\label{source:markov}.
\end{align}
% And we assume that 
% \begin{equation}
% P_{XYZ}(x,y,z)>0\quad \forall (x,y,z)\in \calX\times\calY\times\calZ.\label{source:positive}.
% \end{equation}
Let $P_{X}$, $P_Y$ and $P_Z$ be induced marginal distributions of $P_{XYZ}$. As shown in Figure \ref{systemmodel}, we consider a two-hop hypothesis testing problem with three terminals. The first terminal, which we term the transmitter, observes a source sequence $X^n$ and sends a compression index $M_1$ to the second terminal, which we term the relay. Given $M_1$ and side information $Y^n$, the relay sends another compression index $M_2$ to the third terminal, which we term the receiver. The main task in this problem is to construct hypothesis tests  at both the relay and the receiver to distinguish between
\begin{align}
&\rmH_0:(X^n,Y^n,Z^n)\sim P_{XYZ}^n=P^n_{XY}P^n_{Z|Y},\\*
&\rmH_1:(X^n,Y^n,Z^n)\sim P_{X}^n P_Y^n P_Z^n.
\end{align}

For subsequent analyses, we formally define a code for   hypothesis testing over a two-hop network as follows.
\begin{definition}
\label{HTMH-codedef}
An $(n,N_1,N_2)$-code for   hypothesis testing over a two-hop network consists of 
\begin{itemize}
\item Two encoders:
\begin{align}
f_1&:\calX^n\to \calM_1:=\{1,\ldots,N_1\},\\
f_2&:\calM_1\times \calY^n\to\calM_2:=\{1,\ldots,N_2\},~\mathrm{and}
\end{align}
\item Two decoders
\begin{align}
g_1&:\calM_1\times\calY^n\to \rm\{H_0,H_1\},\\
g_2&:\calM_2\times\calZ^n\to \rm\{H_0,H_1\}.
\end{align}
\end{itemize}
\end{definition}

Given an $(n,N_1,N_2)$-code with encoding and decoding functions $(f_1,f_2,g_1,g_2)$, we define 
acceptance regions for the null hypothesis $\rmH_0$ at the relay and the receiver as
\begin{align}
\calA_{Y,n}&:=\{(m_1,y^n):g_1(m_1,y^n)=\rmH_0\}\label{HTMH-YAcceptregion},\\
\calA_{Z,n}&:=\{(m_2,z^n):g_2(m_2,z^n)=\rmH_0\}\label{HTMH-ZAcceptregion}
\end{align}
respectively. We also define conditional distributions 
\begin{align}
P_{M_1|X^n}(m_1|x^n)&:=1\{f_1(x^n_1)=m_1\},\label{def:pm1gxn}\\*
P_{M_2|Y^nM_1}(m_2|y^n,m_1)&:=1\{f_2(m_1,y^n)=m_2\}\label{def:pm2g}.
\end{align}
Thus, for a $(n,N_1,N_2)$-code characterized by $(f_1,f_2,g_1,g_2)$, the joint distribution of random variables $(X^n,Y^n,Z^n,M_1,M_2)$ under the null hypothesis $\rmH_0$ is given by
\begin{align}
P_{X^nY^nZ^nM_1M_2}(x^n,y^n,z^n,m_1,m_2)
&=P^n_{XYZ}(x^n,y^n,z^n) P_{M_1|X^n}(m_1|x^n)P_{M_2|Y^nM_1}(m_2|y^n,m_1)\label{HTMH-PM0},
\end{align}
and under the alternative hypothesis $\rmH_1$ is given by
\begin{align}
\barP_{X^nY^nZ^nM_1M_2}(x^n,y^n,z^n,m_1,m_2)=P^n_X(x^n)P^n_Y(y^n)P_Z^n(z^n)P_{M_1|X^n}(m_1|x^n)P_{M_2|Y^nM_1}(m_2|y^n,m_1).\label{HTMH-PM1}
\end{align}
Now, let $P_{Y^nM_1}$ and $P_{Z^nM_2}$ be   marginal distributions induced by $P_{X^nY^nZ^nM_1M_2}$ and let $\barP_{Y^nM_1}$ and $\barP_{Z^nM_2}$ be   marginal distributions induced by $\barP_{X^nY^nZ^nM_1M_2}$. Then, we can define the type-I and type-II error probabilities at the relay as
\begin{align}
\beta_1&:=P_{M_1Y^n}(\calA_{Y,n}^\rmc)\label{HTMH-YPtyep1},\\
\beta_2&:=\barP_{M_1Y^n}(\calA_{Y,n})\label{HTMH-YPtyep2}
\end{align}
respectively and at the receiver as
\begin{align}
\eta_1&:=P_{M_2Z^n}(\calA_{Z,n}^\rmc)\label{HTMH-ZPtyep1},\\
\eta_2&:=\barP_{M_2Z^n}(\calA_{Z,n})\label{HTMH-ZPtyep2}
\end{align}
respectively. Clearly, $\beta_1,\beta_2,\eta_1$, and $\eta_2$ are functions of $n$ but we suppress these dependencies for brevity.

Given above definitions, the achievable rate-exponent region for the hypothesis testing problem in a two-hop network is defined as follows.
\begin{definition}
\label{def:region}
Given any $(\varepsilon_1,\varepsilon_2)\in(0,1)^2$, a tuple $(R_1,R_2,E_1,E_2)$ is said to be {\em $(\varepsilon_1,\varepsilon_2)$-achievable} if there exists a sequence of $(n,N_1,N_2)$-codes such that 
\begin{align}
\limsup_{n\to \infty}\frac1{n}\log N_i&\leq R_i,\quad\forall i\in\{1,2\},\\
\limsup_{n\to \infty}\beta_1&\leq \varepsilon_1,\label{T1error}\\
\limsup_{n\to \infty}\eta_1&\leq \varepsilon_2,\label{T2error}\\
\liminf_{n\to \infty}-\frac1{n}\log \beta_2&\geq E_1,\\
\liminf_{n\to \infty}-\frac1{n}\log \eta_2&\geq E_2.
\end{align}
The closure of the set of all $(\varepsilon_1,\varepsilon_2)$-achievable rate-exponent tuples is called the  {\em $(\varepsilon_1,\varepsilon_2)$-rate-exponent region} and is denoted as $\calR(\varepsilon_1,\varepsilon_2)$.
Furthermore, define the {\em rate-exponent region} as
\begin{align}
\calR&:=\calR(0,0).\label{HTMH-Weakdef}
\end{align}
\end{definition}

\subsection{Existing Results}
In the following, we recall the exact characterization of $\calR$ given by Salehkalaibar, Wigger and Wang~\cite[Prop.~2]{salehkalaibar2017hypothesis}. For this purpose, define the following set of joint distributions
\begin{align}\label{HTMH-Qset}
\calQ:=\{Q_{XYZUV}\in\calP(\calX\times\calY\times\calZ\times\calU\times\calV):Q_{XYZ}=P_{XYZ},\ U-X-Y,\ V-Y-Z\}.
\end{align}
 Given $Q_{XYZUV}\in\calQ$, define the following set
\begin{align}\label{HTMH-Cap}
\calR(Q_{XYZUV}):=\big\{(R_1,R_2,E_1,E_2):R_1&\geq I_Q(U;X),R_2\geq I_Q(V;Y),\nn\\*
E_1&\leq I_Q(U;Y),E_2\leq I_Q(U;Y)+I_Q(V;Z)\big\}
\end{align}
Finally, let
\begin{align}
\calR^*&:=\bigcup_{Q_{XYZUV}\in\calQ}\calR(Q_{XYZUV}).
\end{align}

\begin{theorem}
\label{th1:sadaf}
The rate-exponent region $\calR$ for the hypothesis testing  over a two-hop network  problem satisfies
\begin{align}
\calR=\calR^*.
\end{align}
\end{theorem}

In the following, inspired by Oohama's variational characterization of rate regions for multiuser information theory~\cite{oohama2015exponent,oohama2016wynerziv}, we provide an alternative characterization of $\calR^*$. For this purpose, given any $(b,c,d)\in\bbR_+^3$ and any $Q_{XYZUV}\in\calQ$, let
\begin{align}
\rmR_{b,c,d}(Q_{XYZUV})&:=-I_Q(U;Y)+bI_Q(U;X)-c(I_Q(U;Y)+I_Q(V;Z))+dI_Q(V;Y)\label{HTMH-Rbcdq}.
\end{align}
be a linear combination of the mutual information terms in \eqref{HTMH-Cap}. Furthermore, define
\begin{align}
\rmR_{b,c,d}&:=\min_{Q_{XYZUV}\in \calQ}\rmR_{b,c,d}(Q_{XYZUV})\label{HTMH-Rbcd}.
\end{align}
An alternative characterization of $\calR^*$ is given by
\begin{align}\label{HTMH-Region}
\calR^*=\bigcap_{(b,c,d)\in \bbR_+^3}\big\{(R_1,R_2,E_1,E_2):-E_1+bR_1-cE_2+dR_2\geq \rmR_{b,c,d}\big\}.
\end{align}

\section{Strong Converse Theorem}

\subsection{The case $\varepsilon_1+\varepsilon_2<1$}
\begin{theorem}\label{result:TH1}
Given any $(\varepsilon_1,\varepsilon_2)\in(0,1)^2$ such that $\varepsilon_1+\varepsilon_2<1$ and any $(b,c,d)\in\bbR_+^3$, for any $(n,N_1,N_2)$-code such that $\beta_1\leq \varepsilon_1$, $\eta_1\leq \varepsilon_2$, we have
\begin{align}\label{result:TH1EQ}
\log \beta_2+b\log N_1+ c\log \eta_2+d\log N_2 \geq n\rmR_{b,c,d}+\Theta(n^{3/4}\log n).
\end{align}
\end{theorem}
The proof of Theorem \ref{result:TH1} is given in Section \ref{sec:proof}. Several remarks are in order.

First, using the alternative expression of the rate-exponent region in \eqref{HTMH-Region}, we conclude that for any $(\varepsilon_1,\varepsilon_2)\in(0,1)^2$ such that $\varepsilon_1+\varepsilon_2<1$, we have $\calR(\varepsilon_1,\varepsilon_2)=\calR^*$. This result significantly strengthens the weak converse result in \cite[Prop.~2]{salehkalaibar2017hypothesis} in which it was shown that $\calR(0,0)=\calR^*$.

Second, it appears difficult to establish the strong converse result in Theorem \ref{result:TH1} using existing  classical techniques including   image-size characterizations (a consequence of the blowing-up lemma)~\cite{csiszar2011information,tian2008successive} and the perturbation approach~\cite{wei2009strong}. In Section~\ref{sec:proof}, we judiciously combine two recently proposed strong converse techniques by Liu, van Handel, and Verd\'u~\cite{liu2017beyond} and by Tyagi and Watanabe~\cite{tyagi2018strong}. In particular, we use the strong converse technique based on reverse hypercontractivity in \cite{liu2017beyond} to bound the exponent of the type-II error probability at the receiver and the strong converse technique in \cite{tyagi2018strong}, which leverages an appropriate change-of-measure technique and replaces hard Markov constraints with soft information costs,  to analyze the exponent of type-II error probability at the relay. Finally, inspired by the single-letterization steps in \cite[Lemma C.2]{liubeyond} and~\cite{tyagi2018strong}, we single-letterize the derived multi-letter bounds from the previous steps to obtain the desired result in Theorem \ref{result:TH1}.

Third, we briefly comment on the apparent necessity of combining the two techniques in \cite{liu2017beyond} and \cite{tyagi2018strong} instead of applying just one of them to obtain  Theorem \ref{result:TH1}. The first step to apply the technique in \cite{tyagi2018strong} is to construct a ``truncated source distribution'' which is supported on a smaller set (often defined in terms of the decoding region) and is not too far away from the true source distribution in terms of the relative entropy. For our problem, the source satisfies the Markov chain $X^n-Y^n-Z^n$. If we na\"ively apply the techniques in \cite{tyagi2018strong}, the Markovian property would not hold for the truncated source $(\tilX^n,\tilY^n,\tilZ^n)$. On the other hand, it appears rather challenging to extend the techniques in \cite{liu2017beyond} to the hypothesis testing over a multi-hop network  problem since the techniques therein rely heavily on constructing   semi-groups and it is difficult to devise   appropriate forms of such semi-groups to be used and analyzed in this multi-hop setting. Therefore, we carefully combine the two techniques in \cite{liu2017beyond} and \cite{tyagi2018strong} to ameliorate the aforementioned problems. In particular, we first use the technique in \cite{tyagi2018strong} to construct a truncated source $(\tilX^n,\tilY^n)$ and then let the conditional distribution of $\tilZ^n$ given $(\tilX^n,\tilY^n)$ be given by the {\em true} conditional source distribution $P_{Z|Y}^n$ to maintain the Markovian property of the source (see \eqref{HTMH-S1}). Subsequently, in the analysis of error exponents, we   use the technique in \cite{liu2017beyond} to analyze the exponent of type-II error probability at the receiver to circumvent the need to construct new semi-groups. % that are amenable to analysis.

\begin{figure}[t]
\centering
\setlength{\unitlength}{0.5cm}
\scalebox{0.9}{
\begin{picture}(19,6)
\linethickness{1pt}
\put(2,0){\makebox(0,0){$X^n$}}
\put(7,0){\makebox(0,0){$Y^n$}}
\put(12,0){\makebox(0,0){$Z^n$}}
%put encoders and decoder
\put(11,2){\framebox(2,2)}
\put(6,2){\framebox(2,2)}
\put(1,2){\framebox(2,2)}
%put names for encoders and decoder
\put(6.7,2.8){\makebox{$f_2$}}
\put(1.7,2.8){\makebox{$f_1$}}
\put(11.7,2.8){\makebox{$g$}}
%put vectors to encoders and decoder
\put(2,0.5){\vector(0,1){1.5}}
\put(7,0.5){\vector(0,1){1.5}}
\put(12,0.5){\vector(0,1){1.5}}
%put vectors between encoders and decoder
\put(2,4){\line(0,1){2}}
\put(2,6){\line(1,0){10}}
\put(7,6){\vector(0,-1){2}}
\put(12,6){\vector(0,-1){2}}
\put(8,3){\vector(1,0){3}}
\put(9.5,3.5){\makebox(0,0){$M_2$}}
\put(1.5,4.5){\makebox(0,0){$M_1$}}
\put(13,3){\vector(1,0){6}}
\put(15.8,3.5){\makebox(0,0){$\rmH_0:P_{XYZ}^n$}}
\put(16,2.5){\makebox(0,0){$\rmH_1:P_{XY}^nP_Z^n$}}
\end{picture}}
\caption{Cascaded hypothesis testing against independence by Zhao and Lai~\cite{zhao2018distributed}.}
\label{systemmodel2}
\end{figure}
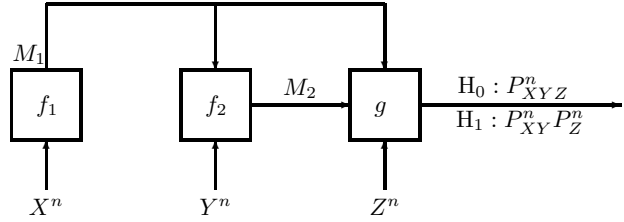

Finally, we remark that the techniques (or a subset of the techniques) used to prove Theorem \ref{result:TH1} can also be used to establish a strong converse result for other multiterminal hypothesis testing against independence problems, e.g., hypothesis testing over the Gray-Wyner network~\cite{wigger2016testing}, the interactive hypothesis testing problem~\cite{xiang2012interactive} and the cascaded hypothesis testing problem~\cite{zhao2018distributed}. In particular, for the testing against independence case in \cite{zhao2018distributed} (as shown in Figure \ref{systemmodel2}), a strong converse result was established by a subset of the present authors in \cite{cao2018itw}.

\subsection{The case $\varepsilon_1+\varepsilon_2>1$}
In this subsection, we consider the case where the sum of type-I error probabilities at the relay and the receiver is upper bounded by a quantity strictly greater than one. For ease of presentation of our results, let
\begin{align}
\calQ_2
&:=\{Q_{XYZU_1U_2V}\in\calQ(\calX\times\calY\times\calZ\times\calU_1\times\calU_2\times\calV):\nn\\
&\quad\quad\quad Q_{XYZ}=P_{XYZ},U_1-X-Y,~U_2-X-Y,~V-Y-Z\}\label{def:calQ2}.
\end{align}
Given any $Q_{XYZU_1U_2V}\in\calQ_2$, define the following set of rate-exponent tuples
\begin{align}\label{HTMH-Cap2}
\tilde{\calR}(Q_{XYZU_1U_2V})
:=\Big\{(R_1,R_2,E_1,E_2):
\nn&R_1\geq \max\{I_Q(U_1;X),I_Q(U_2;X)\},~R_2\geq I_Q(V;X),\\
&E_1\leq I_Q(U_1;Y),~E_2\leq I_Q(U_2;Y)+I_Q(V;Z)
\Big\}.
\end{align}
Furthermore, define
\begin{align}
\tilde{\calR}&:=\bigcup_{Q_{XYZU_1U_2V}}\tilde{\calR}(Q_{XYZU_1U_2V}).
\end{align}
Given any $Q_{XYZU_1U_2V}\in\calQ_2$ and $(b_1,b_2,c,d)\in\bbR_+^4$, define the following linear combination of the mutual information terms
\begin{align}
\tilde{\rmR}_{b_1,b_2,c,d}(Q_{XYZU_1U_2V})
&:=-I_Q(U_1;Y)+b_1I_Q(U_1;X)+b_2I_Q(U_2;X)-c(I_Q(U_2;Y)+I_Q(V;Z))+dI_Q(V;Y),
\end{align}
and let
\begin{align}
\tilde{\rmR}_{b_1,b_2,c,d}
&:=\min_{Q_{XYZU_1U_2V}}\tilde{\rmR}_{b_1,b_2,c,d}(Q_{XYZU_1U_2V}).
\end{align}
Then, based on~\cite{oohama2015exponent,oohama2016wynerziv}, an alternative characterization of $\tilde{\calR}$ is given by
\begin{align}
\tilde{\calR}
&=\bigcup_{(b_1,b_2,c,d)\in\bbR_+^4}\big\{(R_1,R_2,E_1,E_2):
-E_1+b_1R_1+b_2R_1-cE_2+dR_2\geq \tilde{\rmR}_{b_1,b_2,c,d}
\big\}\label{alternative:tildecalR}.
\end{align}

Analogously to Theorem \ref{result:TH1}, we obtain the following result.
\begin{theorem}
\label{result:TH2}
Given any $(\varepsilon_1,\varepsilon_2)\in(0,1)^2$ such that $\varepsilon_1+\varepsilon_2>1$ and any $(b_1,b_2,c,d)\in\bbR_+^4$, for any $(n,N_1,N_2)$-code such that $\beta_1\leq \varepsilon_1$, $\eta_1\leq \varepsilon_2$, we have
\begin{align}\label{result:TH2EQ}
\log \beta_2+b_1\log N_1+ b_2\log N_1+c\log \eta_2+d\log N_2 \geq n\tilde{\rmR}_{b_1,b_2,c,d}+\Theta(n^{3/4}\log n).
\end{align}
\end{theorem}
The proof of Theorem \ref{result:TH2} involves applying the proof of Theorem \ref{result:TH1} to two special cases of the problem in Figure \ref{systemmodel}: i)   hypothesis testing with communication constraint where the receiver does not exist, and ii) the relay is not required to output a decision. Thus, the proof of Theorem \ref{result:TH2} is omitted for brevity.

Using Theorem \ref{result:TH2}, we obtain the following proposition, which together with the first remark of Theorem \ref{result:TH1} provides a strong converse theorem for the  problem of hypothesis testing against independence over a two-hop network  when $\varepsilon_1+\varepsilon_2\neq 1$.
\begin{proposition}
\label{sc:g1}
For any $(\varepsilon_1,\varepsilon_2)\in(0,1)^2$ such that $\varepsilon_1+\varepsilon_2>1$, we have
\begin{align}
\calR(\varepsilon_1,\varepsilon_2)=\tilde{\calR}.
\end{align}
\end{proposition}
The converse proof of Proposition \ref{sc:g1} follows from Theorem \ref{result:TH2} and the alternative characterization of $\tilde{\calR}$ in \eqref{alternative:tildecalR}. The achievability proof is inspired by \cite[Theorem 5]{tian2008successive} and is provided in Appendix \ref{proof:prop}. The main idea is that we can time-share between two close-to optimal coding schemes, each of which corresponds to one special case of the current problem as mentioned after Theorem \ref{result:TH2}.

Finally, we remark that the case when $\varepsilon_1+\varepsilon_2=1$ is not included. See \cite[Sec.\ III.D]{tian2008successive} for a discussion of this subtle case.

%As argued in \cite{tian2008successive}, the reason why we cannot obtain the strong converse result in this case is analogous to the reason why the behavior of separate source-channel coding is unknown when the entropy rate of the memoryless source is exactly equal to the capacity of the noisy channel.

\section{Proof of Theorem \ref{result:TH1}}
\label{sec:proof}
We present the proof of strong converse theorem for the hypothesis testing over the two-hop network in this section. The proof follows by judiciously combing the techniques in \cite{liu2017beyond} and \cite{tyagi2018strong} and is separated into three main steps. First, we construct a truncated source distribution $P_{\tilX^n\tilY^n\tilZ^n}$ and show that this truncated distribution is not too different from $P_{XYZ}^n$ in terms of the relative entropy. Subsequently, we analyze the exponents of type-II error probabilities at the relay and the receiver under the constraint that their type-I error probabilities are non-vanishing. Finally, we single-letterize the constraints on rate and error exponents to obtain desired result in Theorem \ref{result:TH1}.

To begin with, let us fix an $(n,N_1,N_2)$-code with functions $(f_1,f_2,g_1,g_2)$ such that the type-I error probabilities are bounded above by $\varepsilon_1\in(0,1)$ and $\varepsilon_2\in(0,1)$ respectively, i.e., $\beta_1\leq \varepsilon_1$ and $\eta_1\leq \varepsilon_2$.\footnote{We note from \eqref{T1error} and \eqref{T2error} that $\beta_1\leq \varepsilon_1+o(1)$ and $\beta_2\leq \varepsilon_2+o(1)$. Since the $o(1)$ terms are immaterial in the subsequent analyses, they are omitted for brevity.}

\subsection{Construction of A Truncated Distribution}

Paralleling the definitions of acceptance regions in \eqref{HTMH-YAcceptregion} and \eqref{HTMH-ZAcceptregion}, we define the following acceptance regions at the relay and the receiver as
\begin{align}
&\calD_{Y,n}=\{(x^n,y^n):g_1(y^n,f_1(x^n))=\rmH_0\},\label{aregiony}\\
&\calD_{Z,n}=\{(x^n,y^n,z^n):g_2(z^n,f_2(f_1(x^n),y^n))=\rmH_0\}\label{aregionz},
\end{align}
respectively. Note that the only difference between $\calA_{Y,n}$ and $\calD_{Y,n}$ lies in whether we consider the compression index $m_1$ or the original source sequence $x^n$. Recalling the definitions of the type-I error probabilities for the relay denoted by $\beta_1$ in \eqref{HTMH-YPtyep1} and for the receiver denoted by $\eta_1$ in \eqref{HTMH-ZPtyep1}, and using \eqref{aregiony} and  \eqref{aregionz}, we conclude that
\begin{align}
P^n_{XY}(\calD_{Y,n})&=1-\beta_1,\label{aryprob}\\
P^n_{XYZ}(\calD_{Z,n})&=1-\eta_1\label{arzprob}.
\end{align}
For further analysis, given any $m_2\in\calM_2$, define a conditional acceptance region at the receiver (conditioned on $m_2$) as
\begin{align}
\calG(m_2)&:=\{z^n:g_2(z^n,m_2)=\rmH_0\}\label{calG:m2}.
\end{align}
For ease of notation, given any $(x^n,y^n)\in\calX^n\times\calY^n$, we use $\calG(x^n,y^n)$ and $\calG(f_2(f_1(x^n),y^n))$ (here $f_2(f_1(x^n),y^n)$ plays the role of $m_2$ in \eqref{calG:m2}) interchangeably and define the following set
\begin{align}
\calB_n:=\Big\{(x^n,y^n):P_{Z|Y}^n(\calG(x^n,y^n)|y^n)\geq \frac{1-\varepsilon_1-\varepsilon_2}{1+3\varepsilon_2-\varepsilon_1}\Big\}\label{HTMH-B}.
\end{align}
Combining \eqref{aregionz}, \eqref{arzprob} and \eqref{calG:m2}, we obtain
\begin{align}
1-\varepsilon_2&\leq P^n_{XYZ}(\calD_{Z,n})\\
&=\sum_{(x^n,y^n)\in \calB_n}P^n_{XY}(x^n,y^n)P^n_{Z|Y}(\calG(x^n,y^n)|y^n)+\sum_{(x^n,y^n)\not\in \calB_n}P^n_{XY}(x^n,y^n)P^n_{Z|Y}(\calG(x^n,y^n)|y^n)\\
&\leq P^n_{XY}(\calB_n)+(1-P^n_{XY}(\calB_n))\frac{1-\varepsilon_1-\varepsilon_2}{1+3\varepsilon_2-\varepsilon_1}.
\end{align}
Thus, we have
\begin{align}
P^n_{XY}(\calB_n)\geq \frac{3-3\varepsilon_2+\varepsilon_1}{4}\label{lbcalb}.
\end{align}
For subsequent analyses, let
% let \red{what if $\min_y P_Y(y)=0$?} \blue{We have now changed the definition of $\mu$ a bit and this definition is used in \eqref{HTMH-L3PF7}. From the definition of set $\calQ_1$ in \eqref{HTMH-Q1set}, we know that if $P_Y(y)=0$ then definitely $Q_Y^{(\gamma)}(y)=0$ and thus the analysis leading to \eqref{HTMH-L3PF7} follows.}
\begin{align}
\mu&:=\Big(\min_{y:P_Y(y)>0} P_Y(y)\Big)^{-1}\label{HTMH-Defmu},\\
\theta_n&:=\sqrt{\frac{3\mu}{n}\log\frac{8|\calY|}{1-\varepsilon_1-\varepsilon_2}}\label{HTMH-Deftheta},
\end{align} 
and define the typical set $\calT_n(P_Y)$ as
\begin{align}\label{def:calT}
\calT_n(P_Y)=\{y^n: |\hatP_{y^n}(y)-P_Y(y)|\leq \theta_n P_Y(y)\quad \forall y\in \calY\}.
\end{align}
Using the Chernoff bound, we conclude that when $n$ is sufficiently large,
\begin{align}\label{lbtypical}
P_Y^n(\calT_n(P_Y))\geq 1-\frac{1-\varepsilon_1-\varepsilon_2}{4}.
\end{align}
Now, define the following set
\begin{align}
\calC_n&:=\calB_n\cap\calD_{Y,n}\cap(\calX^n\times \calT_n(P_Y))\label{def:calCn}.
\end{align} 
Then, combining \eqref{aryprob}, \eqref{lbcalb} and \eqref{lbtypical}, we conclude that when $n$ is sufficiently large,
\begin{align}\label{HTMH-PC}
P^n_{XY}(\calC_n)\geq 1-P^n_{XY}(\calB_n^\rmc)-P^n_{XY}(\calD_{Y,n}^\rmc)-P_Y^n(\calT_n^\rmc(P_Y))\geq \frac{1-\varepsilon_1-\varepsilon_2}2.
\end{align}

%In the remainder of this subsection, we will define a truncated distribution $P_{\tilX^n\tilY^n\tilZ^n}$ using $\calC_n$ and show that this distribution is not too far away from the true source distribution $P_{XYZ}^n$ in terms of the relative entropy.

Let the truncated distribution $P_{\tilX^n\tilY^n\tilZ^n}$ be defined as
\begin{align}\label{HTMH-S1}
P_{\tilX^n\tilY^n\tilZ^n}(x^n,y^n,z^n)&:=\frac{P^n_{XY}(x^n,y^n)1\{(x^n,y^n)\in\calC_n\}}{P^n_{XY}(\calC_n)}P^n_{Z|Y}(z^n|y^n).
\end{align}
Using the result in \eqref{HTMH-PC}, we have that the marginal distribution $P_{\tilX^n}$ satisfies that for any $x^n\in\calX^N$, 
\begin{align}
P_{\tilX^n}(x^n)
&=\sum_{y^n,z^n}P_{\tilX^n\tilY^n\tilZ^n}(x^n,y^n,z^n)\\
&\leq \frac{P^n_X(x^n)}{P^n_{XY}(\calC_n)}\leq \frac{2P^n_X(x^n)}{1-\varepsilon_1-\varepsilon_2}\label{HTMH-PX}.
\end{align}
Analogously to \eqref{HTMH-PX}, we obtain that
\begin{align}
P_{\tilY^n}(y^n)&\leq \frac{2P^n_Y(y^n)}{1-\varepsilon_1-\varepsilon_2},\quad\forall~y^n\in\calY^n\label{HTMH-PY},\\
P_{\tilZ^n}(z^n)&\leq \frac{2P_Z^n(z^n)}{1-\varepsilon_1-\varepsilon_2},\quad\forall~z^n\in\calZ^n\label{HTMH-PZ}.
\end{align}
Finally, note that
\begin{align}\label{HTMH-Divergence}
D(P_{\tilX^n\tilY^n\tilZ^n}\|P^n_{XYZ})
&=D(P_{\tilX^n\tilY^n}\|P^n_{XY})\\
&=\log\frac1{P^n_{XY}(\calC_n)}\\
&\leq \log\frac2{1-\varepsilon_1-\varepsilon_2}\label{kldsmall}.
\end{align}

\subsection{Analyses of the Error Exponents of Type-II Error Probabilities}
\subsubsection{Type-II error probability $\beta_2$ at the relay}
Let $\tilM_1$ and $\tilM_2$ be the outputs of encoders $f_1$ and $f_2$ respectively when the tuple of source sequences $(\tilX^n,\tilY^n,\tilZ^n)$ is distributed according to $P_{\tilX^n\tilY^n\tilZ^n}$ defined in \eqref{HTMH-S1}. Thus, recalling the definitions in \eqref{def:pm1gxn}, \eqref{def:pm2g} and \eqref{HTMH-S1}, we find that the joint distribution of $(\tilX^n,\tilY^n,\tilZ^n,\tilM_1,\tilM_2)$ is given by
\begin{align}\label{HTMH-JP}
P_{\tilX^n\tilY^n\tilZ^n\tilM_1\tilM_2}(x^n,y^n,z^n,m_1,m_2)=P_{\tilX^n\tilY^n\tilZ^n}(x^n,y^n,z^n)P_{M_1|X^n}(m_1|x^n)P_{M_2|Y^nM_1}(m_2|y^n,m_1).
\end{align}
Let $P_{\tilM_1\tilY^n}$ be induced by $P_{\tilX^n\tilY^n\tilZ^n\tilM_1\tilM_2}$. Combining \eqref{HTMH-YAcceptregion} and \eqref{HTMH-S1}, we conclude that
\begin{align}
P_{\tilM_1\tilY^n}(\calA_{Y,n})
&=\sum_{\substack{x^n,y^n,z^n,m_1,m_2:\\g_1(m_1,y^n)=\rmH_0}}P_{\tilX^n\tilY^n\tilZ^n\tilM_1\tilM_2}(x^n,y^n,z^n,m_1,m_2)\\
&=\sum_{x^n,y^n:g_1(f_1(x^n),y^n)=\rmH_0}\frac{P^n_{XY}(x^n,y^n)1\{(x^n,y^n)\in\calC_n\}}{P^n_{XY}(\calC_n)}\\
&=\sum_{x^n,y^n}\frac{P^n_{XY}(x^n,y^n)1\{(x^n,y^n)\in\calC_n\}}{P^n_{XY}(\calC_n)}\label{Temp1}\\
&=1.
\end{align}
where \eqref{Temp1} follows from the definition of $\calD_{Y,n}$ in \eqref{aregiony} and the fact that $\calD_{Y,n}\subseteq \calC_n$.\par
Thus, using the data processing inequality for the relative entropy and   the definition of $\beta_2$ in \eqref{HTMH-YPtyep2}, we obtain that
\begin{align}
D(P_{\tilM_1\tilY^n}\|P_{M_1}P_Y^n)
&\geq D_\rmb(P_{\tilM_1\tilY^n}(\calA_{Y,n})\|P_{M_1}P_Y^n(\calA_{Y,n}))\\
&=-\log \big(P_{M_1}P_Y^n(\calA_{Y,n})\big)\\
&=-\log \beta_2\label{uppbeta2}.
\end{align}

Furthermore, recalling that $M_1$ denotes the output of encoder $f_1$ when $(X^n,Y^n,Z^n)\sim P_{XYZ}^n$ and $\tilM_1$ denotes the output of encoder $f_1$ when $(X^n,Y^n,Z^n)\sim P_{\tilX^n\tilY^n\tilZ^n}$, and using the result in \eqref{HTMH-PX}, we conclude  that 
\begin{align}
P_{\tilM_1}(m_1)
&=\sum_{x^n,y^n,z^n:f_1(x^n)=m_1}P_{\tilX^n\tilY^n\tilZ^n}(x^n,y^n,z^n)\\
&=\sum_{x^n:f_1(x^n)=m_1}P_{\tilX^n}(x^n)\\
&\leq \sum_{x^n:f_1(x^n)=m_1}\frac{2P_X^n(x^n)}{1-\varepsilon_1-\varepsilon_2}\\
&\leq \frac{2P_{M_1}(m_1)}{1-\varepsilon_1-\varepsilon_2}\label{uppPtilm1},
\end{align}
for any $m_1\in\calM_1$.
Thus, combining \eqref{HTMH-PY}, \eqref{uppbeta2} and \eqref{uppPtilm1}, we have
\begin{align}
-\log\beta_2
&\leq D(P_{\tilM_1\tilY^n}\|P_{M_1}P_Y^n)\\
&=D(P_{\tilM_1\tilY^n}\|P_{\tilM_1}P_{\tilY^n})+\bbE_{P_{\tilM_1\tilY^n}}\left[\log\frac{P_{\tilM_1}(\tilM_1)P_{\tilY^n}(\tilY^n)}{P_{M_1}(\tilM_1)P_Y^n(\tilY^n)}\right]\label{def:calTradeoff100}\\
&\leq D(P_{\tilM_1\tilY^n}\|P_{\tilM_1}P_{\tilY^n})+\bbE_{P_{\tilM_1\tilY^n}}\left[\log\frac{\frac{2P_{M_1}(\tilM_1)}{1-\varepsilon_1-\varepsilon_2}\frac{2P_Y^n(\tilY^n)}{1-\varepsilon_1-\varepsilon_2}}{P_{M_1}(\tilM_1)P_Y^n(\tilY^n)}\right]\label{def:calTradeoff10}\\
&= I(\tilM;\tilY^n)+2\log\frac2{1-\varepsilon_1-\varepsilon_2}.\label{def:calTradeoff1}
\end{align}
%\red{In \eqref{def:calTradeoff10} should the arguments of the prob distributions  $M_1$ be $\tilM_1$ and $Y^n$ be $\tilY^n$? Also please insert these arguments in the probability distributions in the last term in  \eqref{def:calTradeoff100}} \blue{Thanks. We have now made the change.}
\subsubsection{Type-II error probability $\eta_2$ at the receiver}
In this subsection, we analyze the error exponent of the type-II error probability at the receiver. For this purpose, we make use of the method introduced in \cite{liu2017beyond} based on reverse hypercontractivity. We define the following additional notation:
\begin{itemize}
\item Give $P_{YZ}\in\calP(\calY\times\calZ)$, define\footnote{In the subsequent analysis, we only consider the case  $\alpha>1$. When $\alpha=1$, choosing $t=\frac{1}{\sqrt{n}}$ instead of the choice in \eqref{chooseT}, we can obtain a similar upper bound for $-\log \eta_2$ as in \eqref{result1}, where the only difference is that $\Psi(n,\varepsilon_1,\varepsilon_2)$ should be replaced by another term scaling in order $\Theta(\sqrt{n})$.}
\begin{align}
\alpha:=\max_{y,z}\frac{P_{Z|Y}(z|y)}{P_Z(z)}\in (1,\infty).
\end{align}
%\red{As I said, we should discuss the case $\alpha=1$. At least in a footnote or small remark.} \blue{Yes. We have now added a footnote to discuss the case when $\alpha=1$. Most proof follows and we just need to refine a parameter $t$.}
\item Given any $(\varepsilon_1,\varepsilon_2)\in(0,1)^2$ such that $\varepsilon_1+\varepsilon_2<1$, let
\begin{align}
\Psi(n,\varepsilon_1,\varepsilon_2):=2\sqrt{n(\alpha-1)\log\frac{1+3\varepsilon_2-\varepsilon_1}{1-\varepsilon_1-\varepsilon_2}}.
\end{align}
\item Give any $m_2\in\calM_2$ and $z^n\in\calZ^n$, let
\begin{align}
h(m_2,z^n):=1\{z^n\in \calG(m_2)\}.
\end{align}
\item Two operators in~\cite[Eqns.~(25), (26), (29)]{liu2017beyond}
\begin{align}
\Lambda_{\alpha,t}&=(\exp(-t)+\alpha(1-\exp(-t))P_Z)^{\otimes n}\label{op1},\\
T_{y^n,t}&=\prod_{i=1}^n (\exp(-t)+(1-\exp(-t))P_{Z|y_{i}})\label{op2}.
\end{align}
\end{itemize}
Note that in \eqref{op2}, we use the convenient notation $P_{Z|y}(z)=P_{Z|Y}(z|y)$. The two operators in \eqref{op1} and \eqref{op2} will be used to lower bound $D(P_{\tilZ^n\tilM_2}\|P_Z^n\barP_{M_2})$ via a variational formula of the relative entropy (cf.\ \cite[Section 4]{liu2017beyond}).

Let $P_{\tilZ^n\tilM_2}$, $P_{\tilZ^n|\tilM_2}$, $P_{\tilZ^n|\tilY^n}$ be induced by the joint distribution $P_{\tilX^n\tilY^n\tilZ^n\tilM_1\tilM_2}$ in \eqref{HTMH-JP} and let $\barP_{M_2}$ be induced by the joint distribution $\barP_{X^nY^nZ^nM_1M_2}$ in \eqref{HTMH-PM1}. Invoking the variational formula for the relative entropy \cite[Eqn.~(2.4.67)]{raginsky2013concentration} and recalling the notation $P(f)=\bbE_P[f]$, we have
\begin{align}\label{HTMH-VFD}
D(P_{\tilZ^n\tilM_2}\|P_Z^n\barP_{M_2})
&\geq P_{\tilZ^n\tilM_2} \big(\log \Lambda_{\alpha,t}h(\tilM_2,\tilZ^n)\big)-\log \big((P_Z^n\barP_{M_2})\big(\Lambda_{\alpha,t}h(M_2,Z^n)\big)\big).
\end{align}
Given any $m_2\in\calM_2$, similar to \cite[Eqns.~(18)--(21)]{liu2017beyond}, we obtain
\begin{align}
&P_Z^n(\Lambda_{\alpha,t}h(m_2,Z^n))\nn\\
&=P_Z^n\big((\exp(-t)+\alpha(1-\exp(-t))P_Z)^{\otimes n}h(m_2,Z^n)\big)\\
&=\big(\exp(-t)+\alpha(1-\exp(-t))\big)^n P_Z^n \big(h(m_2,Z^n)\big)\\
&\leq \exp((\alpha-1)nt)P_Z^n \big(h(m_2,Z^n)\big)\label{liu18-21}.
\end{align}
Thus, averaging over $m_2$ with distribution $\barP_{M_2}$ on both sides of \eqref{liu18-21}, we have
\begin{align}
&(P_Z^n\barP_{M_2})(\Lambda_{\alpha,t}h(M_2,Z^n))\nn\\
%&=\barP_{M_2}\Big(P_Z^n(\Lambda_{\alpha,t}h(M_2,Z^n))\Big)\\
&\leq \exp((\alpha-1)nt) (\barP_{M_2}P_Z^n)\big(h(M_2,Z^n)\big)\\
&=\exp((\alpha-1)nt) \eta_2\label{HTMH-PA1},
\end{align}
where \eqref{HTMH-PA1} follows from the definition of $\eta_2$ in \eqref{HTMH-ZPtyep2}.

Furthermore, given any $\tilm_2\in\calM_2$, we obtain
\begin{align}
&P_{\tilZ^n|\tilm_2}(\log \Lambda_{\alpha,t}h(\tilm_2,\tilZ^n))\\
&=\Big(\sum_{\tily^n}P_{\tilZ^n|\tily^n}P_{\tilY^n|\tilM_2}(\tily^n|\tilm_2)\Big)(\log \Lambda_{\alpha,t}h(\tilm_2,\tilZ^n))\\
&=\sum_{\tily^n}P_{\tilY^n|\tilM_2}(\tily^n|\tilm_2) P_{\tilZ^n|\tily^n}(\log \Lambda_{\alpha,t}h(\tilm_2,\tilZ^n))\\
&\geq\sum_{\tily^n}P_{\tilY^n|\tilM_2}(\tily^n|\tilm_2) P_{\tilZ^n|\tily^n}(\log T_{y^n,t}h(\tilm_2,\tilZ^n))\label{HTMH-PA2}\\
&\geq \sum_{\tily^n}P_{\tilY^n|\tilM_2}(\tily^n|\tilm_2) \left(1+\frac1{t}\right)\log P_{\tilZ^n|\tily^n}\big(h(\tilm_2,\tilZ^n)\big)\label{HTMH-PA3}\\
&=\left(1+\frac1{t}\right)\Big(\sum_{\tily^n}P_{\tilY^n|\tilM_2}(\tily^n|\tilm_2)\log P_{ \tilZ^n|\tily^n}(\calG(\tilm_2))\Big)\label{HTMH-PA4}.
\end{align}
where \eqref{HTMH-PA2} follows from \cite[Lemma 4]{liu2017beyond} and \eqref{HTMH-PA3} follows similarly to \cite[Eqns.~(14)-(17)]{liu2017beyond}.

Thus, averaging on both sides of \eqref{HTMH-PA4} over $\tilm_2$ with distribution $P_{\tilM_2}$ and using the definition of the joint distribution $P_{\tilX^n\tilY^n\tilZ^n\tilM_1\tilM_2}$ in \eqref{HTMH-JP}, we obtain that
\begin{align}
\nn&P_{\tilZ^n\tilM_2}(\log \Lambda_{\alpha,t}h(\tilM_2,\tilZ^n))\\*
&\geq \left(1+\frac1{t}\right)\Big(\sum_{\tily^n,\tilm_2}P_{\tilY^n\tilM_2}(\tily^n,\tilm_2)\log P_{ \tilZ^n|\tily^n}(\calG(\tilm_2))\Big)\\
&=\left(1+\frac1{t}\right)\sum_{\tilx^n,\tily^n,\tilm_1,\tilm_2}\Bigg(P_{\tilX^n\tilY^n}(\tilx^n,\tily^n)1\{\tilm_1=f_1(\tilx^n),\tilm_2=f_2(\tilm_1,\tily^n)\}\log \bigg(\sum_{\tilz^n:g_2(\tilz^n,\tilm_2)=\rmH_0}P_{Z|Y}^n(\tilz^n|\tily^n)\bigg)\Bigg)\\
&=\left(1+\frac1{t}\right)\bigg(\sum_{\substack{\tilx^n,\tily^n}}\frac{P^n_{XY}(\tilx^n,\tily^n)1\{(\tilx^n,\tily^n)\in\calC_n\}}{P^n_{XY}(\calC_n)}\log P_{Z|Y}^n(\calG(\tilx^n,\tily^n)|\tily^n)\bigg)\\
&\geq \left(1+\frac1{t}\right)\log \frac{1-\varepsilon_1-\varepsilon_2}{1+3\varepsilon_2-\varepsilon_1} ,\label{HTMH-PA5}
\end{align}
where \eqref{HTMH-PA5} follows from the definitions of $\calB_n$ in \eqref{HTMH-B} and $\calC_n$ in \eqref{def:calCn}.

Therefore, combining \eqref{HTMH-VFD}, \eqref{HTMH-PA1} and \eqref{HTMH-PA5} and choosing % $t$ such that
\begin{align}\label{chooseT}
t=\sqrt{\frac{1}{ n(\alpha-1)} \log   \frac{1+3\varepsilon_2-\varepsilon_1}{1-\varepsilon_1-\varepsilon_2}},
\end{align}
via simple algebra, we obtain that
\begin{align}
-\log \eta_2
&\leq D(P_{\tilZ^n\tilM_2}\|P_{Z^n}\barP_{M_2})+\Psi(n,\varepsilon_1,\varepsilon_2)-\log \frac{1-\varepsilon_1-\varepsilon_2}{1+3\varepsilon_2-\varepsilon_1}\label{result1}.
\end{align}

In the following, we further upper bound $D(P_{\tilZ^n\tilM_2}\|P_{Z^n}\barP_{M_2})$. For this purpose, define the following distribution
\begin{align}
\barP_{\tilM_2}(m_2)
&:=\sum_{y^n,m_1}P_{\tilM_1}(m_1)P_{\tilY^n}(y^n)1\{m_2=f_2(m_1,y^n)\}.\label{def:barptilm2}
\end{align}
Combining the results in \eqref{HTMH-PY} and \eqref{uppPtilm1}, and recalling that $\barP_{M_2}$ is induced by joint distribution $\barP_{X^nY^nZ^nM_1M_2}$ in \eqref{HTMH-PM1}, for any $m_2\in\calM_2$, we have
\begin{align}
\barP_{\tilM_2}(m_2)
&\leq \Big(\frac2{1-\varepsilon_1-\varepsilon_2}\Big)^2\bigg(\sum_{y^n,m_1}P_{M_1}(m_1)P_Y^n(y^n)1\{m_2=f_2(f_1(x^n),y^n)\}\bigg)\\
&=\frac{4\barP_{M_2}(m_2)}{(1-\varepsilon_1-\varepsilon_2)^2}\label{uppptilm2}.
\end{align}
Thus, combining \eqref{HTMH-PZ} and \eqref{uppptilm2}, we have
\begin{align}
&D(P_{\tilZ^n\tilM_2}\|P_Z^n\barP_{M_2})\nn\\
&=D(P_{\tilZ^n\tilM_2}\|P_{\tilZ^n}\barP_{\tilM_2})+\bbE_{P_{\tilZ^n\tilM_2}}\bigg[\log\frac{P_{\tilZ^n}(\tilZ^n)\barP_{\tilM_2}(\tilM_2)}{P_Z^n(\tilZ^n)\barP_{M_2}(\tilM_2)}\bigg]\\
&\leq D(P_{\tilZ^n\tilM_2}\|P_{\tilZ^n}\barP_{\tilM_2})+\bbE_{P_{\tilZ^n\tilM_2}}\bigg[\log\frac{\frac{2P_Z^n(\tilZ^n)}{1-\varepsilon_1-\varepsilon_2}\frac{4\barP_{M_2}(\tilM_2)}{(1-\varepsilon_1-\varepsilon_2)^2}}{P_Z^n(\tilZ^n)\barP_{M_2}(\tilM_2)}\bigg]\\
&=D(P_{\tilZ^n\tilM_2}\|P_{\tilZ^n}\barP_{\tilM_2})+3\log \frac2{1-\varepsilon_1-\varepsilon_2}\label{result2}.
\end{align}
%\red{Same comment applies as that after \eqref{def:calTradeoff1}} \blue{Thanks. Have made the change.}
Therefore, combining \eqref{result1} and \eqref{result2}, we have
\begin{align}
-\log \eta_2
&\leq D(P_{\tilZ^n\tilM_2}\|P_{\tilZ^n} \barP_{\tilM_2})+\Psi(n,\varepsilon_1,\varepsilon_2)-\log \frac{1-\varepsilon_1-\varepsilon_2}{1+3\varepsilon_2-\varepsilon_1}-3\log \frac{1-\varepsilon_1-\varepsilon_2}2\label{def:calTradeoff2}.
\end{align}

\subsection{Analyses of Communication Constraints and Single-Letterization Steps}

For any $(n,N_1,N_2)$-code, since $\tilM_i\in\calM_i$ for $i\in\{1,2\}$, we have that
\begin{align}
\log N_1&\geq H(\tilM_1)\geq I(\tilM_1;\tilX^n\tilY^n),\label{HTMH-Rate1}\\
\log N_2&\geq H(\tilM_2)\geq I(\tilM_2;\tilY^n)\label{HTMH-Rate2}.
\end{align}
Furthermore, from the problem setting (see \eqref{HTMH-JP}), we have
\begin{align}
I(\tilM_1;\tilY^n|\tilX^n)=0,\label{HTMH-Zeroterm}
\end{align}
For subsequent analyses, given any $(b,c,d,\gamma)\in\bbR_+^4$, define 
\begin{align}
\rmR_{b,c,d,\gamma}^{(n)}\nn&:=-I(\tilM_1;\tilY^n) +bI(\tilM_1;\tilX^n\tilY^n)-cD(P_{\tilZ^n\tilM_2}\|P_{\tilZ^n}\barP_{\tilM_2})+dI(\tilM_2;\tilY^n)+\gamma I(\tilM_1;\tilY^n|\tilX^n)\nn\\*
&\qquad +(b+d+\gamma)D(P_{\tilX^n\tilY^n}\|P_{X^nY^n})\label{HTMH-Rbcdgn}.
\end{align}
Combining the results in \eqref{kldsmall}, \eqref{def:calTradeoff1}, \eqref{def:calTradeoff2} to \eqref{HTMH-Zeroterm}, for any $\gamma\in\bbR_+$, we obtain
\begin{align}
&\log \beta_2+b\log N_1+c\log \eta_2+d\log N_2+c\Psi(n,\varepsilon_1,\varepsilon_2)\nn\\
&\geq  \rmR_{b,c,d,\gamma}^{(n)}+\log \frac{1-\varepsilon_1-\varepsilon_2}{1+3\varepsilon_2-\varepsilon_1}+(b+d+\gamma+5)\log \frac{1-\varepsilon_1-\varepsilon_2}2\label{nletterresult}.
\end{align}

The proof of Theorem \ref{result:TH1} is complete by the two following lemmas which provide a single-letterized lower bound for $\rmR_{b,c,d,\gamma}^{(n)}$ and relate the derived lower bound to $\rmR_{b,c,d}$. For this purpose, recalling the definition of $\theta_n$ in \eqref{HTMH-Deftheta}, we define the following set of joint distributions
\begin{align}
\calQ_1 &:=\big\{Q_{XYZUV}\in\calP(\calX\times\calY\times\calZ\times\calU\times\calV): \nn\\*
&\qquad Q_{Z|Y}=P_{Z|Y},X-Y-Z,~V-Y-Z, \nn\\*
&\qquad |Q_Y(y)-P_Y(y)|\leq \theta_n P_Y(y),\quad\forall y\in\calY\big\}\label{HTMH-Q1set}.
\end{align}
Given $Q_{XYZUV}\in\calQ_1$, define
\begin{align}
\Delta_{b,d,\gamma}(Q_{XYZUV})&:=(b+\gamma)D(Q_{XY}\|P_{XY})+dD(Q_Y\|P_Y)+\gamma I_Q(U;Y|X)\label{HTMH-Deltabdg}.
\end{align}
Recall the definition of $\rmR_{b,c,d}(Q_{XYZUV})$ in \eqref{HTMH-Rbcdq}. Define
\begin{align}
\rmR_{b,c,d,\gamma}&:=\min_{Q_{XYZUV}\in \calQ_1}\big(\rmR_{b,c,d}(Q_{XYZUV})+\Delta_{b,d,\gamma}(Q_{XYZUV})\big)\label{HTMH-DR1}.
\end{align}

The following lemma presents a single-letterized lower bound for $\rmR_{b,c,d,\gamma}^{(n)}$.
\begin{lemma}\label{HTMH-lemma2}
For any $(b,c,d,\gamma)\in\bbR^4_+$,
\begin{align}
\rmR_{b,c,d,\gamma}^{(n)}\geq n\rmR_{b,c,d,\gamma}.
\end{align}
\end{lemma}
The proof of Lemma \ref{HTMH-lemma2} is inspired by \cite[Prop.~2]{tyagi2018strong} and provided in Appendix \ref{proof:lemma2}.

Combining the results in \eqref{nletterresult} and Lemma \ref{HTMH-lemma2}, we obtain the desired result and this completes the proof of Theorem \ref{result:TH1}.
\begin{lemma}\label{HTMH-lemma3}
Choosing $\gamma=\sqrt{n}$, we have
\begin{align}
n\rmR_{b,c,d,\gamma}+\log \frac{1-\varepsilon_1-\varepsilon_2}{1+3\varepsilon_2-\varepsilon_1}+(b+d+\gamma+5)\log \frac{1-\varepsilon_1-\varepsilon_2}{2}\geq n \rmR_{b,c,d}+\Theta(n^{3/4}\log n).
\end{align}
\end{lemma}
The proof of Lemma \ref{HTMH-lemma3} is inspired by \cite[Lemma C.2]{liubeyond} and provided in Appendix \ref{proof:lemma3}.

% \begin{lemma}
% \label{HTMH-lemma1}
% For any $(b,c,d)\in\bbR_+^3$ and any $(\varepsilon_1,\varepsilon_2)$ satisfying $\varepsilon_1+\varepsilon_2<1$, choosing $\gamma=\sqrt{n}$, we have
% \begin{align}
% \rmR_{b,c,d,\gamma}^{(n)}+\log \frac{1-\varepsilon_1-\varepsilon_2}{1+3\varepsilon_2-\varepsilon_1}+(b+d+\gamma+5)\log \frac{1-\varepsilon_1-\varepsilon_2}2\geq n\rmR_{b,c,d}+\Theta(n^{3/4}\log n).
% \end{align}
% \end{lemma}
% The proof of Lemma \ref{HTMH-lemma1} is given in Appendix \ref{HTMH-prooflemma1}. In the proof of Lemma \ref{HTMH-lemma1}, we first single-letterize the multi-letter expression of $\rmR_{b,c,d,\gamma}^{(n)}$ (see \eqref{HTMH-Rbcdgn}) and then relate the single-letterized formula to $\rmR_{b,c,d}$ (see \eqref{HTMH-Rbcd}).

\section{Discussion and Future Work}
We strengthened the result in \cite[Prop.~2]{salehkalaibar2017hypothesis} by deriving a strong converse theorem for hypothesis testing against independence over a two-hop network with communication constraints (see Figure \ref{systemmodel}). In our proof, we  judiciously combined two recently proposed strong converse techniques \cite{liu2017beyond,tyagi2018strong}. The apparent necessity of doing so comes from the Markovian requirement in the source distribution (recall \eqref{source:markov}) and is reflected in the construction of a truncated distribution in \eqref{HTMH-S1} to ensure the Markovian structure of the source sequences   is preserved. Subsequently, due to this constraint, the application the strong converse technique by Tyagi and Watanabe in \cite{tyagi2018strong} was only amenable in  analyzing the type-II error exponent at the relay. On the other hand, to analyze the type-II error exponent at the receiver, we need to carefully adapt the strong converse technique based on reverse hypercontractivity by Liu, van Handel and Verd\'u in \cite{liu2017beyond}. Furthermore, to complete the proof, we carefully combine the single-letterization techniques in \cite{liu2017beyond,tyagi2018strong}.
 
Another important take-home message is the techniques (or a subset of the techniques) used in this paper can be applied to strengthen the results of other multiterminal hypothesis testing   against independence problems. If the source distribution has no Markov structure, it is usually the case that one can directly apply the technique by Tyagi and Watanabe~\cite{tyagi2018strong} to obtain strong converse theorems. Such examples include \cite{xiang2012interactive,wigger2016testing,zhao2016distributed}. On the other hand, if the source sequences admit Markovian structure, then it appears necessary to combine techniques in~\cite{liu2017beyond,tyagi2018strong} to obtain strong converse theorems, just as it was done in this paper. 

Finally, we discuss some avenues for future research. In this paper, we only derived the strong converse but not a second-order converse result as was done in \cite[Section 4.4]{liu2017beyond} for the problem of hypothesis testing  against independence with a communication constraint~\cite{ahlswede1986hypothesis}. Thus, in the future, one may refine the proof in the current paper by deriving second-order converse or exact second-order asymptotics. Furthermore, one may also consider deriving strong converse theorems or simplifying existing strong converse proofs for hypothesis testing problems with both communication and privacy constraints such as that in \cite{gilani2018distributed} by using the techniques   in the current paper. It  is also interesting to explore whether the current techniques can be applied to obtain strong converse theorems for  hypothesis testing with zero-rate  compression problems~\cite{shalaby1992multiterminal}.

\appendix
\subsection{Achievability Proof of Proposition \ref{sc:g1}}
\label{proof:prop}
Fix any joint distribution $Q_{XYZU_1U_2V}\in\calQ_2$. Let $(f_1',g_1')$ be an encoder-decoder  pair with rate $R_1=I_Q(U_1;X)$ for the hypothesis testing  with communication constraint problem~\cite{ahlswede1986hypothesis} (i.e., no receiver in Figure \ref{systemmodel}) such that the type-II error probability decays exponentially fast at speed no smaller than $E_1=I_Q(U_1;Y)$ and the type-I error probability is vanishing, i.e., $\log N_1'\leq nR_1$, $\beta_2'\leq \exp(-nE_1)$ and $\beta_1'\leq \varepsilon_1'$ for any $\varepsilon_1'>0$. Furthermore, let $(f_1'',f_2'',g_1'',g_2'')$ be a tuple of encoders and decoders with rates $(R_1,R_2)=(I_Q(U_2;X),I_Q(V;Y))$ for the problem in Figure \ref{systemmodel} such that the type-II error probability at the receiver decays exponentially fast at speed no smaller $E_2=I_Q(V;Z)$ and type-I error probability at the receiver is vanishing, i.e., $\log N_1''\leq nR_1$, $\log N_2''\leq nR_2$, $\eta_2''\leq \exp(-nE_2)$ and $\eta_1''\leq \varepsilon_2'$ for any $\varepsilon_2'>0$. Such tuples of encoders and decoders exist as proved in \cite{ahlswede1986hypothesis} and \cite{salehkalaibar2017hypothesis}. Furthermore, let $\calA_1'\subseteq\calX^n\times\calY^n$ be the acceptance region associated with $(f_1',g_1')$ at the relay and let $\calA_2'\subseteq\calX^n\times\calY^n\times\calZ^n$ be the acceptance region associated with $(f_1'',f_2'',g_1'',g_2'')$ at the receiver.

Now, let us partition the source space $\calX^n$ into two disjoint sets $\calX_1^n$ and $\calX_2^n$ such that $\calX_1^n\cup\calX_2^n=\calX^n$, $P_X^n(\calX_1^n)>1-\varepsilon_1$ and $P_X^n(\calX_2^n)>1-\varepsilon_2$. We construct an $(n,N_1,N_2)$-code as follows. Given a source sequence $X^n$, if $X^n\in\calX_1^n$, then encoder $f_1'$ is used; and if otherwise, the encoder $f_1''$ is used. Furthermore, an additional bit indicating whether $X^n\in\calX_1^n$ is also sent to the relay and further forwarded to the receiver by the relay. Given encoded index $M_1$, if $X^n\in\calX_1^n$, the relay uses decoder $g_1'$ to make the decision; otherwise, if $X^n\in\calX_2^n$, the relay declares hypothesis $\rmH_1$ to be true. Furthermore, in both cases, the relay transmits an index $M_2$ using encoder $f_2''$. Given the index $M_2$, if $X^n\in\calX_1^n$, the receiver declares hypothesis $\rmH_1$ to be true; otherwise, the receiver uses decoder $g_2''$ to make the decision.

The performance of the constructed $(n,N_1,N_2)$-code is as follows. In terms of rates, we have
\begin{align}
\log N_1\leq nR_1+1,\\
\log N_2\leq nR_2+1.
\end{align}
The type-I error probability at the relay satisfies that
\begin{align}
1-\beta_1
&=P_{XY}^n\{\calA_1'\cap(\calX_1^n\times\calY^n)\}\\
&\geq P_X^n\{\calX_1^n\}-P_{XY}^n\{(\calA_1')^{\rmc}\}\\
&\geq 1-\varepsilon_1\label{large},
\end{align}
where \eqref{large} follows when $n$ is sufficiently large and thus $\varepsilon_1'$ can be made arbitrarily close to zero. Furthermore, the type-II error probability at the relay can be upper bounded as follows
\begin{align}
\beta_2
&=P_X^nP_Y^n\{\calA_1'\cap(\calX_1^n\times\calY^n)\}\\
&\leq P_X^nP_Y^n\{\calA_1'\}\\
&=\beta_2'\\
&\leq \exp(-nE_1').
\end{align}
Similarly, for $n$ sufficiently large, the error probabilities at the receiver can be upper bounded as follows
\begin{align}
\eta_1
&=1-P_{XYZ}^n\{\calA_2''\cap(\calX_2^n\times\calY^n\times\calZ^n)\}\\
&\leq 1-P_X^n(\calX_2^n)+P_{XYZ}^n\big((\calA_2')^\rmc\big)\\
&\leq \varepsilon_2,
\end{align}
and
\begin{align}
\eta_2
&=P_X^nP_Y^nP_Z^n\{\calA_2''\cap(\calX_2^n\times\calY^n\times\calZ^n)\}\\
&\leq P_X^nP_Y^nP_Z^n\{\calA_2''\}\\
&\leq \exp(-nE_2'').
\end{align}
The achievability proof of Proposition \ref{sc:g1} is now complete.

\subsection{Proof of Lemma \ref{HTMH-lemma2}}
\label{proof:lemma2}
Recall the definition of distribution $\barP_{\tilM_2}$ (see \eqref{def:barptilm2}). Noting that $P_{\tilM_2}$ is the marginal distribution induced by $P_{\tilX^n\tilY^n\tilZ^n\tilM_1\tilM_2}$ (see \eqref{HTMH-JP}), we have that for any $\tilm_2\in\calM_2$
\begin{align}
P_{\tilM_2}(\tilm_2)
% &=\sum_{x^n,y^n,z^n,m_1}P_{\tilX^n\tilY^n\tilZ^n}(x^n,y^n,z^n)1\{m_1=f_1(x^n),\tilm_2=f_2(m_1,y^n)\}\\
&=\sum_{y^n,m_1}P_{\tilY^n\tilM_1}(y^n,m_1)1\{\tilm_2=f_2(m_1,y^n)\}.
\end{align}
Thus, applying the data processing inequality for the relative entropy, we have that
\begin{align}
I(\tilM_1;\tilY^n)
&=D(P_{\tilY^n\tilM_1}\|P_{\tilY^n}P_{\tilM_1})\\
&\geq D(P_{\tilM_2}\|\barP_{\tilM_2})\label{dpikl}.
\end{align}
Using \eqref{dpikl} and following similar steps to the proof of weak converse in \cite[Eq.~(186)]{salehkalaibar2017hypothesis}, we obtain
\begin{align}
D(P_{\tilZ^n\tilM_2}\| P_{\tilZ^n}\barP_{\tilM_2})
&=I(\tilM_2;\tilZ^n)+D(P_{\tilM_2}\|\barP_{\tilM_2})\\
&\leq I(\tilM_2;\tilZ^n)+I(\tilM_1;\tilY^n)\label{reversedpi}.
\end{align}
Using \eqref{reversedpi} and the definition of $\rmR_{b,c,d,\gamma}^{(n)}$ in \eqref{HTMH-Rbcdgn}, we have the following lower bound for $\rmR_{b,c,d,\gamma}^{(n)}$
\begin{align}
\rmR_{b,c,d,\gamma}^{(n)}
&\geq-I(\tilM_1;\tilY^n) +b\big(D(P_{\tilX^n\tilY^n}\|P_{X^nY^n})+H(\tilX^n\tilY^n)-H(\tilX^n\tilY^n|\tilM_1)\big)\nn\\
&\quad-c\big(I(\tilM_2;Z^n)+I(\tilM_1;\tilY^n)\big)+d\big(D(P_{\tilX^n\tilY^n}\|P_{X^nY^n})+H(\tilY^n)-h(\tilY^n|\tilM_2)\big)\nn\\
&\quad+\gamma\big(D(P_{\tilX^n\tilY^n}\|P_{X^nY^n})+H(\tilY^n|\tilX^n)-H(\tilY^n|\tilX^n\tilM_1)\big)\label{HTMH-L2PF1}.
\end{align}
The rest of the proof concerns single-letterizing each term in \eqref{HTMH-L2PF1}. For this purpose, for each $j\in[n]$, we define two auxiliary random variables $U_j:=(\tilM_1,\tilX^{j-1},\tilY^{j-1})$ and $V_j:=(\tilM_2,\tilY^{j-1})$ and let $J$ be a random variable which is distributed uniformly over the set $[n]$ and is independent of all other random variables. 

Using standard single-letterization techniques as in \cite{el2011network}, we obtain
\begin{align}
I(\tilM_1;\tilY^n)
&=\sum_{j\in[n]}I(\tilM_1;\tilY_j|\tilY^{j-1})\\
&\leq \sum_{j\in[n]}I(\tilM_1,\tilY^{j-1};\tilY_j)\\
&\leq \sum_{j\in[n]}I(\tilM_1,\tilX^{j-1},\tilY^{j-1};\tilY_j)\\
&=nI(U_J,J;\tilY_J),
\end{align}
and
\begin{align}
H(\tilX^n\tilY^n|\tilM_1)
&=nH(\tilX_J\tilY_J|U_J,J).
\end{align}
Furthermore, analogous to \cite[Prop.~1]{tyagi2018strong}, we obtain that 
\begin{align}
H(\tilX^n\tilY^n)+D(P_{\tilX^n\tilY^n}\|P_{XY}^n)
&=\sum_{x^n,y^n}P_{\tilX^n\tilY^n}(x^n,y^n)\log\frac1{P_{XY}^n(x^n,y^n)}\\
&=\sum_{x^n,y^n}P_{\tilX^n\tilY^n}(x^n,y^n)\sum_{j\in[n]}\log\frac1{P_{XY}(x_j,y_j)}\\
&=\sum_{j\in[n]}P_{\tilX_j\tilY_j}(x_j,y_j)\log\frac1{P_{XY}(x_j,y_j)}\\
&=n \big(H(\tilX_J,\tilY_J)+D(P_{\tilX_JY_J}\|P_{XY})\big).
\end{align}

Subsequently, we can single-letterize $I(\tilM_2;\tilZ^n)$ as follows:
\begin{align}
I(\tilM_2;\tilZ^n)
&=\sum_{j\in[n]}I(\tilM_2;\tilZ_j|\tilZ^{j-1})\\
&\leq \sum_{j\in[n]}I(\tilM_2\tilZ^{j-1}\tilY^{j-1};\tilZ_j)\\
&=\sum_{j\in[n]}I(\tilM_2\tilY^{j-1};\tilZ_j)\label{HTMH-L2PF2}\\
&=nI(V_J,J;\tilZ_J)\label{HTMH-L2PF3},
\end{align}
where \eqref{HTMH-L2PF2} follows from the Markov chain $\tilZ^{j-1}-\tilM_2\tilY^{j-1}-\tilZ_j$ implied by the joint distribution of $(\tilX^n,\tilY^n,\tilZ^n,\tilM_1,\tilM_2)$ in \eqref{HTMH-JP}. Furthermore, using similar proof techniques to \cite[Prop.~1]{tyagi2018strong} and standard single-letterization techniques (e.g.,~in~\cite{csiszar2011information} or~\cite{el2011network}), we obtain that 
\begin{align}
H(\tilY^n|\tilde{X}^n)+D(P_{\tilX^n\tilY^n}\|P_{XY}^n)
&\geq  n \big(H(\tilY_{J}|\tilde{X}_J)+D(P_{\tilX_J\tilY_J}\|P_{XY})\big),\\
H(\tilY^n)+D(P_{\tilX^n\tilY^n}\|P_{XY}^n)
&\geq n \big(H(\tilY_J)+D(P_{Y_J}\|P_Y)\big),\\
H(\tilY^n|\tilM_2)&=nH(\tilY_J|V_J,J),\\
H(\tilY^n|\tilM_1\tilX^n)&\leq nH(\tilY_J|X_J,U_J,J)\label{proof:end}.
\end{align}
Let $U:=(U_J,J)$, $V:=(V_J,J)$, $X':=\tilX_J$, $Y':=\tilY_{J}$ and $Z':=\tilZ_J$. Using the joint distribution $P_{\tilX^n\tilY^n\tilZ^n\tilM_1\tilM_2}$ in \eqref{HTMH-JP}, we conclude that the joint distribution of random variables $(X',Y',Z',U,V)$, denoted by $Q_{X'Y'Z'UV}$,   %(\red{Is ``for $(X',Y',Z',U,V)$'' necessary?} \blue{Yes. It is necessary here since we define $Q_{X'Y'Z'UV}$ as the joint distribution of $(X',Y',Z',U,V)$. It was not defined anywhere. We have paraphrased the sentence a bit to remove confusion.}) 
belongs to the set $\calQ_1$ defined in \eqref{HTMH-Q1set}. The proof of Lemma \ref{HTMH-lemma2} is complete by combining \eqref{HTMH-L2PF1} to \eqref{proof:end} and noting that $I_{Q}(X',Y';U)\geq I_{Q}(X';U)$.

%the proof of final lemma
\subsection{Proof of Lemma \ref{HTMH-lemma3}}
\label{proof:lemma3}
Given any $\gamma\in\bbR_+$, let $Q^{(\gamma)}_{XYZUV}$ achieve the minimum in \eqref{HTMH-DR1}. Recall the definition of $\theta_n$ in \eqref{HTMH-Deftheta} and define a new alphabet $\tilde{\calV}:=\calV\cup \{v^*\}$. We then define a joint distribution $P_{Y\tilV}^{(\gamma)}$ by specifying the  following (conditional) marginal distributions
\begin{align}
P^{(\gamma)}_{\tilV}(v)
&:=\frac1{1+\theta_n}Q^{(\gamma)}_V(v)1\{v\neq v^*\}+\frac{\theta_n}{1+\theta_n}1\{v=v^*\}\label{def:Pgtilv},\\
P^{(\gamma)}_{Y|\tilV}(y|v)
&:=Q^{(\gamma)}_{Y|V}(y|v)1\{v\ne v^*\}  +  \left(\frac{1+\theta_n}{\theta_n}P_Y(y)-\frac1{\theta_n}Q^{(\gamma)}_Y(y) \right)  1\{v=v^*\}. \label{def:pgygtilv}
%%,\quad \forall v\in\calV,\\
%P^{(\gamma)}_{Y|\tilV}(y|v^*)
%&:=\frac{1+\theta_n}{\theta_n}P_Y(y)-\frac1{\theta_n}Q^{(\gamma)}_Y(y)\label{def:pgygtilv}.
\end{align}
Thus,  the induced marginal distribution $P_Y^{(\gamma)}$ satisfies 
\begin{align}
P_Y^{(\gamma)}(y)
&=\sum_{v\in\tilde{\calV}} P_{\tilV}^{(\gamma)}(v)P_{Y|\tilV}^{(\gamma)}(y|v)\\
&=\left(\sum_{v\in\calV}\frac{1}{1+\theta_n}Q_V^{(\gamma)}(v)Q_{Y|V}^{(\gamma)}(y|v)\right)+\left(P_Y(y)-\frac1{1+\theta_n}Q_Y^{(\gamma)}(y)\right)\\
&=P_Y(y)\label{pyg=py}.
\end{align}
Furthermore, let $P^{(\gamma)}_{\tilV|Y}$ be induced by $P^{(\gamma)}_{Y\tilV}$ and define the following distribution
\begin{align}
P^{(\gamma)}_{XYZU\tilV}=P_{XYZ}Q^{(\gamma)}_{U|X}P^{(\gamma)}_{\tilV|Y}\label{HTMH-L3PF1}.
\end{align}
Recall the definition of $\rmR_{b,c,d}(\cdot)$ in \eqref{HTMH-Rbcdq}. The following lemma lower bounds the difference between $\rmR_{b,c,d}(Q^{(\gamma)}_{XYZUV})$  and $\rmR_{b,c,d}(P^{(\gamma)}_{XYZU\tilV})$ and is critical in the proof of Lemma \ref{HTMH-lemma3}.
\begin{lemma}
\label{lemma:afinal}
When $\gamma=\sqrt{n}$, we have
\begin{align}
\rmR_{b,c,d}(Q^{(\gamma)}_{XYZUV})-\rmR_{b,c,d}(P^{(\gamma)}_{XYZU\tilV})\geq \Theta\left(\frac{\log n}{n^{1/4}}\right)\label{almostfinal}.
\end{align}
\end{lemma}
The proof of Lemma \ref{lemma:afinal} is deferred to Appendix \ref{proof:afinal}.

Now, using the assumption that $Q^{(\gamma)}_{XYZUV}$ is a minimizer for $\rmR_{b,c,d,\gamma}$ in \eqref{HTMH-DR1}, the fact that $\Delta_{b,d,\gamma}(Q^{(\gamma)}_{XYZUV})\geq 0$ (see~\eqref{HTMH-Deltabdg}) and the result in \eqref{almostfinal}, we conclude that when $\gamma=\sqrt{n}$,
\begin{align}
\rmR_{b,c,d,\gamma}
&=\rmR_{b,c,d}(Q^{(\gamma)}_{XYZUV})+\Delta_{b,d,\gamma}(Q^{(\gamma)}_{XYZUV})\\
&\geq \rmR_{b,c,d}(P^{(\gamma)}_{XYZU\tilV})+\Theta\left(\frac{\log n}{n^{1/4}}\right)\\
&\geq \rmR_{b,c,d}+\Theta\left(\frac{\log n}{n^{1/4}}\right)\label{finalresult},
\end{align}
where \eqref{finalresult} follows from the definition of $\rmR_{b,c,d}$ in \eqref{HTMH-Rbcd} and the fact that $P^{(\gamma)}_{XYZU\tilV}\in\calQ$ (see \eqref{HTMH-Qset}).

The proof of Lemma \ref{HTMH-lemma3} is complete by using \eqref{finalresult} and noting that when $\gamma=\sqrt{n}$,
\begin{align}
\log \frac{1-\varepsilon_1-\varepsilon_2}{1+3\varepsilon_2-\varepsilon_1}+(b+d+\gamma+5)\log \frac{1-\varepsilon_1-\varepsilon_2}{2}=\Theta(\sqrt{n}).
\end{align}

\subsection{Proof of Lemma \ref{lemma:afinal}}
\label{proof:afinal}
In subsequent analyses, all distributions indicated by $P^{(\gamma)}$ are induced by $P^{(\gamma)}_{XYZU\tilV}$. We have
\begin{align}
D(Q^{(\gamma)}_{XYU}\|P^{(\gamma)}_{XYU})=D(Q^{(\gamma)}_{XY}\|P^{(\gamma)}_{XY})+I_{Q^{(\gamma)}}(U;Y|X).
\end{align}
Recalling the definitions of $\rmR_{b,c,d}$ in \eqref{HTMH-Rbcd} and $\rmR_{b,c,d,\gamma}$ in \eqref{HTMH-DR1}, we conclude that for any $\gamma\in\bbR_+$,
\begin{align}
\rmR_{b,c,d,\gamma}\leq \rmR_{b,c,d}\leq b\log|\calX|+d\log|\calY|=:a'.
\end{align}
Using the definition of $\Delta_{b,d,\gamma}(Q_{XYZUV})$ in \eqref{HTMH-Deltabdg} and recalling that $Q^{(\gamma)}_{XYZUV}$ is a minimizer for $\rmR_{b,c,d,\gamma}$, we have
\begin{align}
\gamma D(Q^{(\gamma)}_{XYU}\|P^{(\gamma)}_{XYU})
&\leq \Delta_{b,d,\gamma}(Q^{(\gamma)}_{XYZUV})\\
&=\rmR_{b,c,d,\gamma}-\rmR_{b,c,d}(Q^{(\gamma)}_{XYZUV})\\
&\leq a'+(c+1)\log|\calY|+c\log|\calZ|=:a\label{uppdqpgamma}.
\end{align}
We can now upper bound $I_{P^{\gamma}}(\tilV;Y)$ as follows:
\begin{align}
I_{P^{(\gamma)}}(\tilV;Y)&=D(P^{(\gamma)}_{Y|\tilV}\|P^{(\gamma)}_Y|P^{(\gamma)}_{\tilV})\\
&=D(P^{(\gamma)}_{Y|\tilV}\|P_Y|P^{(\gamma)}_{\tilV})\label{usepy=}\\
&=\frac1{1+\theta_n}D(Q^{(\gamma)}_{Y|V}\|P_Y|Q^{(\gamma)}_V)+\frac{\theta_n}{1+\theta_n}D\left(\frac{1+\theta_n}{\theta_n}P_Y-\frac1{\theta_n}Q^{(\gamma)}_Y\middle\|P_Y\right)\\
&=\frac1{1+\theta_n}\big(D(Q^{(\gamma)}_{Y|V}\|Q^{(\gamma)}_Y|Q^{(\gamma)}_V)+D(Q^{(\gamma)}_Y\|P_Y)\big)+\frac{\theta_n}{1+\theta_n}D\left(\frac{1+\theta_n}{\theta_n}P_Y-\frac1{\theta_n}Q^{(\gamma)}_Y\middle\|P_Y\right)\\
&\leq \frac1{1+\theta_n}I_{Q^{(\gamma)}}(V;Y)+\frac1{1+\theta_n}\frac{a}{\gamma} +\frac{\theta_n}{1+\theta_n} \log \mu,\label{HTMH-L3PF7}
\end{align}
where \eqref{usepy=} follows from \eqref{pyg=py}, and \eqref{HTMH-L3PF7} follows from the result in \eqref{uppdqpgamma}, the fact that $D(Q^{(\gamma)}_Y\|P_Y)\leq D(Q^{(\gamma)}_{XYU}\|P^{(\gamma)}_{XYU})$ and the definition of $\mu$ in \eqref{HTMH-Defmu}. Thus, when $\gamma=\sqrt{n}$, recalling the definition  of $\theta_n$ in \eqref{HTMH-Deftheta}, we have
\begin{align}
I_{Q^{(\gamma)}}(V;Y)
&\geq I_{P^{(\gamma)}}(\tilV;Y)- \frac{a}{\gamma} -\theta_n \log\mu\\
&=I_{P^{(\gamma)}}(\tilV;Y)+\Theta\left(\frac1{\sqrt{n}}\right)\label{HTMH-L3PF2}.
\end{align}

Similar to \eqref{HTMH-L3PF7}, we obtain
\begin{align}
I_{P^{(\gamma)}}(\tilV;Z)&=D(P^{(\gamma)}_{Z|\tilV}\|P^{(\gamma)}_{Z}|P^{(\gamma)}_{\tilV})\\
&=D(P^{(\gamma)}_{Z|\tilV}\|P_{Z}|P^{(\gamma)}_{\tilV})\\
&=\frac1{1+\theta_n}D(Q^{(\gamma)}_{Z|V}\|P_{Z}|Q^{(\gamma)}_V)+\frac{\theta_n}{1+\theta_n}D\left(\frac{1+\theta_n}{\theta_n}P_Z-\frac1{\theta_n}Q^{(\gamma)}_Z\middle\|P_Z\right)\label{usemarkovq1}\\
&=\frac1{1+\theta_n}\big(D(Q^{(\gamma)}_{Z|V}\|Q^{(\gamma)}_{Z}|Q^{(\gamma)}_V)+D(Q^{(\gamma)}_{Z}\|P_{Z})\big)+\frac{\theta_n}{1+\theta_n}D\left(\frac{1+\theta_n}{\theta_n}P_Z-\frac1{\theta_n}Q^{(\gamma)}_Z
\middle\|P_Z\right)\\
&\geq \frac1{1+\theta_n} I_{Q^{(\gamma)}}(V;Z)\label{HTMH-L3PF4},
\end{align}
where \eqref{usemarkovq1} follows since $Q^{(\gamma)}\in\calQ_1$ (see \eqref{HTMH-Q1set}) implies that $Q^{(\gamma)}_{Z|Y}=P_{Z|Y}$ and the Markov chains $Z-Y-X$ and $V-Y-Z$ holds and thus using \eqref{def:Pgtilv} to \eqref{def:pgygtilv}, we have
\begin{align}
P_{Z|\tilV}^{(\gamma)}(z|v)
&=\frac{\sum_{y}P_{Z|Y}(z|y)P_{\tilV}(v)P_{Y|\tilV}(y|v)}{P_{\tilV}(v)}\\
&=\frac{\sum_y Q^{(\gamma)}_{Z|Y}(z|y)Q^{(\gamma)}_V(v)Q^{(\gamma)}_{Y|V}(y|v)}{Q^{(\gamma)}_V(v)}\\
&=Q^{(\gamma)}_{Z|V}(z|v),%~\forall~v\in\calV,
\end{align}
and
\begin{align}
P_{Z|\tilV}^{(\gamma)}(z|v^*)
&=\frac{\sum_{y}P_{Z|Y}(z|y)P_{\tilV}(v^*)P_{Y|\tilV}(y|v^*)}{P_{\tilV}(v^*)}\\
&=\sum_y Q^{(\gamma)}_{Z|Y}(z|y)\Big(\frac{1+\theta_n}{\theta_n}P_Y(y)-\frac{1}{\theta_n}Q^{(\gamma)}_Y(y)\Big)\\
&=\frac{1+\theta_n}{\theta_n}P_Z(z)-\frac{1}{\theta_n}Q^{(\gamma)}_Z(z),
\end{align}

Therefore, we have
\begin{align}
I_{Q^{(\gamma)}}(V;Z) &\leq (1+\theta_n)I_{P^{(\gamma)}}(\tilV;Z)\\
&\leq I_{P^{(\gamma)}}(\tilV;Z)+\theta_n\log|\calZ|\\
&=I_{P^{(\gamma)}}(\tilV;Z)+\Theta\left(\frac1{\sqrt{n}}\right)\label{HTMH-L3PF3}.
\end{align}

Let $\| P-Q\|$ be the $\ell_1$ norm between $P$ and $Q$ regarded as vectors. Using Pinsker's inequality, the result in \eqref{uppptilm2}, and the data processing inequality for the relative entropy~\cite{cover2012elements}, we obtain
\begin{align}
\|Q^{(\gamma)}_{UX}-P^{(\gamma)}_{UX}\|
&\leq \sqrt{2\log 2 \cdot D(Q^{(\gamma)}_{UX}\|P^{(\gamma)}_{UX})}\label{HTCE-L3PF9}\\
&\leq \sqrt{2\log 2 \cdot D(Q^{(\gamma)}_{XYU}\|P^{(\gamma)}_{XYU})}\label{HTCE-L3PF10}\\
&\leq \sqrt{\frac{2a\log 2}{\gamma}}.
\end{align}

From the support lemma~\cite[Appendix~C]{el2011network}, we conclude that the cardinality of $U$ can be upper bounded by a function depending only on $|\calX|$, $|\calY|$ and $|\calZ|$ (these alphabets are all finite). Thus, when $\gamma=\sqrt{n}$, invoking \cite[Lemma 2.2.7]{csiszar2011information}, we have
\begin{align}
|H(Q^{(\gamma)}_{UX})-H(P^{(\gamma)}_{UX})|\leq  \sqrt{\frac{2a\log 2}{\gamma}}\log\frac{|\calU||\calX|}{\sqrt{\frac{2a\log 2}{\gamma}}}=\Theta\left(\frac{\log n}{n^{1/4}}\right)\label{csiszerlemma}.
\end{align}
Similar to \eqref{csiszerlemma}, we have 
\begin{align}
|I_{Q^{(\gamma)}}(U;X)-I_{P^{(\gamma)}}(U;X)|
&\leq \Theta\left(\frac{\log n}{n^{1/4}}\right)\label{HTMH-L3PF5},\\
|I_{Q^{(\gamma)}}(U;Y)-I_{P^{(\gamma)}}(U;Y)|
&\leq \Theta\left(\frac{\log n}{n^{1/4}}\right)\label{HTMH-L3PF6}.
\end{align}
Combining \eqref{HTMH-L3PF2}, \eqref{HTMH-L3PF3}, \eqref{HTMH-L3PF5} and \eqref{HTMH-L3PF6}, when $\gamma=\sqrt{n}$, using the definition of $\rmR_{b,c,d}(\cdot)$ in \eqref{HTMH-Rbcdq}, we have
\begin{align}
\rmR_{b,c,d}(Q^{(\gamma)}_{XYZUV})
&\geq-(c+1)I_{Q^{(\gamma)}}(U;Y)+bI_{Q^{(\gamma)}}(U;X)-cI_{Q^{(\gamma)}}(V;Z)+dI_{Q_{(\gamma)}}(V;Y)\\
&\geq -(c+1)I_{P^{(\gamma)}}(U;Y)+bI_{P^{(\gamma)}}(U;X)-cI_{P^{(\gamma)}}(\tilV;Z)+dI_{P^{(\gamma)}}(\tilV;Y)+\Theta\left(\frac{\log n}{n^{1/4}}\right)\\
&=\rmR_{b,c,d}(P^{(\gamma)}_{XYZU\tilV})+\Theta\left(\frac{\log n}{n^{1/4}}\right).
\end{align}
The proof of Lemma \ref{lemma:afinal} is now complete.

%\red{Reference [12] seems incomplete. Please fix all references and make them ``uniform''} \blue{Thanks. We have now cited ISIT version in most places and only mention original [12] now [18] very few times. \cite{liubeyond} was not published nor uploaded to arxiv. It was a bit hard to cite it but we need to use Lemma C.2 in the paper for singe-letterization in the proof of Lemmas \ref{HTMH-lemma3} and \ref{lemma:afinal}.}
\section*{Acknowledgments} 
The authors acknowledge Dr. Sadaf Salehkalaibar for drawing our attention to \cite[Prop.~2]{salehkalaibar2017hypothesis} and providing helpful comments and suggestions.

Daming Cao is supported by the China Scholarship Council with No.\ 201706090064 and the National Natural Science Foundation of China under Grant 61571122. Lin Zhou is supported by NUS RSB grants (C-261-000-207-532 and C-261-000-005-001).

\bibliographystyle{IEEEtran}
\bibliography{MAC}
\end{document}

%% file: preamble.tex
\usepackage[mathscr]{eucal}
\usepackage[cmex10]{amsmath}
\usepackage{epsfig,epsf,psfrag}
\usepackage{amssymb,amsmath,amsthm,amsfonts,latexsym}
\usepackage{amsmath,graphicx,bm,xcolor,url,overpic}
\usepackage{fixltx2e}%ordering of single and double column floats
\usepackage{array}%array and tabular environments
\usepackage{verbatim}
\usepackage{bm}
\usepackage{algorithmic}
\usepackage{algorithm}
\usepackage{verbatim}
\usepackage{textcomp}
\usepackage{mathrsfs}
\usepackage{epstopdf}

\newcommand{\openone}{\leavevmode\hbox{\small1\normalsize\kern-.33em1}}

\catcode`~=11 \def\UrlSpecials{\do\~{\kern -.15em\lower .7ex\hbox{~}\kern .04em}} \catcode`~=13 

\allowdisplaybreaks[4]

\newcommand{\nn}{\nonumber}

\newcommand{\calA}{\mathcal{A}}
\newcommand{\calB}{\mathcal{B}}
\newcommand{\calC}{\mathcal{C}}
\newcommand{\calD}{\mathcal{D}}

\newcommand{\calG}{\mathcal{G}}

\newcommand{\calM}{\mathcal{M}}

\newcommand{\calP}{\mathcal{P}}
\newcommand{\calQ}{\mathcal{Q}}
\newcommand{\calR}{\mathcal{R}}

\newcommand{\calT}{\mathcal{T}}
\newcommand{\calU}{\mathcal{U}}
\newcommand{\calV}{\mathcal{V}}

\newcommand{\calX}{\mathcal{X}}
\newcommand{\calY}{\mathcal{Y}}
\newcommand{\calZ}{\mathcal{Z}}

\newcommand{\rmb}{\mathrm{b}}

\newcommand{\rmc}{\mathrm{c}}

\newcommand{\rmH}{\mathrm{H}}

\newcommand{\rmR}{\mathrm{R}}

\newcommand{\bbE}{\mathsf{E}}

\newcommand{\bbN}{\mathbb{N}}

\newcommand{\bbR}{\mathbb{R}}

\DeclareMathAlphabet{\mathbsf}{OT1}{cmss}{bx}{n}
\DeclareMathAlphabet{\mathssf}{OT1}{cmss}{m}{sl}% slanted sans serif

\DeclareSymbolFont{bsfletters}{OT1}{cmss}{bx}{n}  
\DeclareSymbolFont{ssfletters}{OT1}{cmss}{m}{n}
\DeclareMathSymbol{\bsfGamma}{0}{bsfletters}{'000}
\DeclareMathSymbol{\ssfGamma}{0}{ssfletters}{'000}
\DeclareMathSymbol{\bsfDelta}{0}{bsfletters}{'001}
\DeclareMathSymbol{\ssfDelta}{0}{ssfletters}{'001}
\DeclareMathSymbol{\bsfTheta}{0}{bsfletters}{'002}
\DeclareMathSymbol{\ssfTheta}{0}{ssfletters}{'002}
\DeclareMathSymbol{\bsfLambda}{0}{bsfletters}{'003}
\DeclareMathSymbol{\ssfLambda}{0}{ssfletters}{'003}
\DeclareMathSymbol{\bsfXi}{0}{bsfletters}{'004}
\DeclareMathSymbol{\ssfXi}{0}{ssfletters}{'004}
\DeclareMathSymbol{\bsfPi}{0}{bsfletters}{'005}
\DeclareMathSymbol{\ssfPi}{0}{ssfletters}{'005}
\DeclareMathSymbol{\bsfSigma}{0}{bsfletters}{'006}
\DeclareMathSymbol{\ssfSigma}{0}{ssfletters}{'006}
\DeclareMathSymbol{\bsfUpsilon}{0}{bsfletters}{'007}
\DeclareMathSymbol{\ssfUpsilon}{0}{ssfletters}{'007}
\DeclareMathSymbol{\bsfPhi}{0}{bsfletters}{'010}
\DeclareMathSymbol{\ssfPhi}{0}{ssfletters}{'010}
\DeclareMathSymbol{\bsfPsi}{0}{bsfletters}{'011}
\DeclareMathSymbol{\ssfPsi}{0}{ssfletters}{'011}
\DeclareMathSymbol{\bsfOmega}{0}{bsfletters}{'012}
\DeclareMathSymbol{\ssfOmega}{0}{ssfletters}{'012}

\newcommand{\hatH}{\hat{H}}

\newcommand{\tilm}{\tilde{m}}
\newcommand{\tilM}{\tilde{M}}

\newcommand{\hatP}{\hat{P}}

\newcommand{\tilV}{\tilde{V}}

\newcommand{\tilx}{\tilde{x}}
\newcommand{\tilX}{\tilde{X}}

\newcommand{\tily}{\tilde{y}}
\newcommand{\tilY}{\tilde{Y}}

\newcommand{\tilz}{\tilde{z}}
\newcommand{\tilZ}{\tilde{Z}}

\newcommand{\barP}{\bar{P}}

\newtheorem{theorem}{Theorem} 
\newtheorem{lemma}[theorem]{Lemma}

\newtheorem{proposition}[theorem]{Proposition}

\newtheorem{definition}{Definition}